\documentclass[journal,onecolumn]{IEEEtran}

\usepackage{cite}
\usepackage[cmex10]{amsmath}
\usepackage{amssymb}
\usepackage{amsfonts}
\usepackage{array}
\usepackage{verbatim}
\usepackage{tikz}
\usepackage{microtype}
\usepackage{enumerate}
\usepackage{mathtools}
\usepackage{float}

\newcommand{\al}{\mathfrak{a}}
\newcommand{\bl}{\mathfrak{b}}
\newcommand{\cl}{\mathfrak{c}}
\newcommand{\dl}{\mathfrak{d}}
\newcommand{\I}{\mathcal{I}1}
\newcommand{\II}{\mathcal{I}2}
\newcommand{\III}{\mathcal{I}3}
\newcommand{\IV}{\mathcal{I}4}
\newcommand{\V}{\mathcal{I}5}
\newcommand{\VI}{\mathcal{I}6}

\newcommand{\ie}{\,\emph{i.e.},\;}
\newcommand{\prob}{\mathrm{Prob}}
\newcommand{\C}{\mathrm{C}}
\newcommand{\info}{\mathrm{I}}
\newcommand{\h}{\mathit{H}}
\newcommand{\const}{\mathit{\theta}}
\newcommand{\word}[1]{\mathrm{#1}} 

\newcommand{\Ra}{\Rightarrow}

\newtheorem{theorem}{Theorem}
\newtheorem{definition}{Definition}
\newtheorem{lemma}{Lemma}
\newtheorem{corollary}{Corollary}
\newtheorem{proposition}{Proposition}

\newtheorem{example}{Example}
\newtheorem{remark}{Remark}

\newcommand*\lon{%
       \mskip1mu
        \relax
        {:}%
        \mskip1mu
        \relax
}

\begin{document}

\title{Conditional Information Inequalities for Entropic and Almost Entropic Points
\thanks{
This paper is accepted to the IEEE Transactions on Information Theory
(DOI 10.1109/TIT.2013.2274614).
Results of this paper were presented in part at IEEE~ISIT~2011~\cite{condineq} and 
IEEE~ITW~2012~\cite{kr-itw12}; also some results of this paper were previously published in preprint~\cite{kr-arxiv11}.
This work was supported in part by  NAFIT ANR-08-EMER-008-01 and EMC ANR-09-BLAN-0164-01 grants.}
}

\author{Tarik Kaced\thanks{T.~Kaced is with the Institute of Network Coding at
    the Chinese University of Hong Kong, Shatin.}  and 
Andrei Romashchenko\thanks{A.~Romashchenko is with LIRMM, CNRS \& Univ. Montpellier II, on leave from IITP of the Russian Academy of Sciences, Moscow.}
}

\maketitle

\begin{abstract}

We study conditional linear information inequalities, \ie linear inequalities for Shannon 
entropy that hold for distributions whose joint entropies meet some linear constraints. 
We prove that some conditional  information inequalities cannot be extended
to any unconditional linear inequalities. 
Some of these conditional inequalities  hold for almost entropic points, while
others do not. We also discuss some counterparts of conditional information inequalities
for Kolmogorov complexity.

\end{abstract}



\IEEEpeerreviewmaketitle



\section{Introduction}


In this paper we consider  discrete random variables with a finite range.  
Let $(X_1,\dotsc,X_n)$ be  a joint distribution of $n$ random variables.
Then several standard information quantities are defined for this distribution.
The basic  quantities are the Shannon entropies of each variable 
$\h(X_i)$, the entropies of  pairs $\h(X_i,X_j)$, the entropies of triples,
quadruples, and so on. 
The other standard  information quantities are the conditional entropies
$\h(X_i|X_j)$, the mutual informations  $I(X_i\lon X_j)$, and 
the conditional mutual informations $I(X_i\lon X_j|X_k)$. All these
quantities can be represented as linear combinations 
of the plain Shannon entropies: 
$$
\begin{array}{ccl}
\h(X_i|X_j)&=&\h(X_i,X_j)-\h(X_j),\\ 
I(X_i\lon X_j) &=& \h(X_i)+\h(X_j)-\h(X_i,X_j),\\ 
I(X_i\lon X_j|X_k)&=&\h(X_i,X_k)+\h(X_j,X_k)
  -\h(X_i,X_j,X_k)-\h(X_k).
\end{array}
$$
Thus, these entropies values 
determine all other standard information quantities of the distribution.

Although there is no linear dependence between 
different entropies  $\h(X_{i_1},\dotsc,X_{i_l})$, their values cannot be arbitrary,
they must match some constraints.  The most important constraints for entropies
(of jointly distributed variables) can be written as linear inequalities.
For instance, it is well known that  every distribution satisfies
$$
\h(X_i,X_j)\ge \h(X_i)
$$
(which means that $\h(X_j|X_i)\ge 0$),
 $$
 \h(X_i)+\h(X_j)\ge \h(X_i,X_j) 
 $$
(which can be expressed   in the standard notation as $I(X_i\lon X_j)\ge 0$), and
 $$
 \h(X_i,X_k)+\h(X_j,X_k) \ge \h(X_i,X_j,X_k) + \h(X_k) 
 $$
(\ie  $I(X_i\lon X_j|X_k)\ge 0$ in the standard notation). 
Notice how the first two inequalities are special cases of the third. 
Basic inequalities extend back to Shannon's original papers on information theory.
We can obtain other instances of the basic inequality if we substitute the individual variables
$X_i,X_j,X_k$ by any tuples of variables. Also, any linear combination of
these basic inequalities with positive coefficients is again a valid inequality. 
The inequalities for entropies that can be obtained in this way
(combinations of instances of the basic inequalities summed up with some 
positive coefficients) are referred to as \emph{Shannon-type inequalities}, see,
e.g., \cite{yeung-book}.

Linear inequalities for entropies are ``general laws'' of information which
are widely used in information theory  to describe fundamental limits in information transmission,  
compression,  secrecy, etc.  Information inequalities  have also applications beyond 
information theory, e.g., in combinatorics and in group theory, see the survey 
in \cite{some-survey}.  So, it is natural to ask whether there exist linear inequalities 
for entropy that hold for all distributions but are not Shannon-type.
N.~Pippenger asked in 1986 a general question \cite{Pippenger} : what is
the class of all linear inequalities for Shannon entropy?  

The first example of a non-Shannon-type inequality was discovered by Z.~Zhang and
R.W.~Yeung in 1998. Quite many (in fact, \emph{infinitely} many)
other linear inequalities for entropy have been found since then. It was discovered 
that  the set of all valid linear inequalities for entropy cannot be reduced to any
finite number of inequalities: F.~Mat\'{u}\v{s} proved that for
$n\ge 4$  the cone of all linear inequalities for $n$-tuples of random
variables is not polyhedral \cite{Matus-inf}. There is still no full answer to
Pippenger's question -- a simple characterization of the class of all
linear inequalities for Shannon entropy is yet to be found.

In this paper we investigate a slightly different class of ``laws of the
information theory''.  We consider \emph{conditional} information inequalities,
\ie linear inequalities for entropies that hold only for distributions that
meet some linear constraints on entropies. To explain this notion we start
with very basic examples. 

\smallskip 

\begin{example}\label{ex:1} If $I(A\lon B)=0$, then $\h(A)+\h(B)\le \h(A,B)$. This follows immediately
from the definition of the mutual information.
\end{example}

\smallskip 

\begin{example}\label{ex:2}  If $I(A\lon B)=0$, then $ \h(A)+\h(B)+\h(C)\le \h(A,C)+\h(B,C).$
This follows from an unconditional Shannon-type  inequality: for all $A,B,C$
$$
\h(A)+\h(B)+\h(C)\le \h(A,C)+\h(B,C)+I(A\lon B).
$$
(This inequality is equivalent to the sum of two basic inequalities: $\h(C|A,B)\ge 0$ and 
$I(A\lon B|C)\ge 0$.)
\end{example}

\smallskip

\begin{example} \label{ex:3}
If  $I(E\lon C|D)=I(E\lon D|C)=I(C\lon D|E)=0$, then 
$$
I(C\lon D) \le I(C\lon D|A)+ I(C\lon D|B)+I(A\lon B).
$$
This is a corollary of the non-Shannon-type inequality from~\cite{MMRV}
(a slight generalization of the inequality proven by Zhang and Yeung in~\cite{ZY98}),
\begin{IEEEeqnarray*}{rCl}
I(C\lon D) &\le& I(C\lon D|A)+ I(C\lon D|B)+I(A\lon B) + \\ &&\: + 
[I(E\lon C|D)+I(E\lon D|C)+I(C\lon D|E)],
\end{IEEEeqnarray*}
which holds for all $(A,B,C,D,E)$ (without any constraints on the distribution).
\end{example}


These three examples are quite different from each other, but they follow the same template:
\medskip
\begin{quote}\it
If \textup[linear equalities hold for entropies of $(A,B,C,\ldots)$\textup] 
then \textup[a linear inequality holds for these $(A,B,C,\ldots)$\textup].
\end{quote}
\medskip
This is what we call a \emph{conditional information inequality} 
(also referred to as a \emph{constrained information inequality}).

In Examples~\ref{ex:1}--\ref{ex:3} the  conditional inequalities are directly derived 
 from the corresponding unconditional ones. But not all
proofs of conditional inequalities are that straightforward.
In 1997 Z.~Zhang and R.W.~Yeung
came up with a conditional inequality
\begin{eqnarray}\label{zy97}
\label{eq-star}
\begin{array}{l}
\mbox{if }I(A\lon B)=I(A\lon B|C)=0,\mbox{ then }   
I(C\lon D) \le I(C\lon D|A)+ I(C\lon D|B)+I(A\lon B), 
\end{array}
\end{eqnarray}
see \cite{ZY97}.
If we wanted to prove~(\ref{eq-star}) similarly to Examples~\ref{ex:1}--\ref{ex:3}
above, then we should first prove an unconditional inequality
\begin{eqnarray}\label{eq-star2}
\begin{aligned}
I(C\lon D) \le I(C\lon D|A)+ I(C\lon D|B)+I(A\lon B) 
+\lambda_1I(A\lon B)+\lambda_2 I(A\lon B|C)
\end{aligned}
\end{eqnarray}
with some ``Lagrange multipliers'' $\lambda_1,\lambda_2\ge 0$.  
However, the proof  in \cite{ZY97} does not follow this  scheme.  
Can we find an alternative proof of~(\ref{eq-star})
that would be based on an instance of~(\ref{eq-star2}), for some 
$\lambda_1$ and $\lambda_2$? 
In this paper we show that this is impossible. We prove that, 
whatever the values $\lambda_1,\lambda_2$, unconditional 
Inequality~(\ref{eq-star2})  does not hold for Shannon entropy.

Note that  in \cite{ZY97} it was already proven that Inequality~(\ref{eq-star}) cannot
be deduced from Shannon-type inequalities (the only linear inequalities
for  entropy known by 1997, when \cite{ZY97} was published). We prove a stronger statement:
(\ref{eq-star}) cannot be deduced directly from  any unconditional
linear inequalities for entropy (Shannon-type or non-Shannon-type, 
known or yet unknown).

Since conditional Inequality~(\ref{eq-star}) cannot be extended
to any unconditional Inequality~(\ref{eq-star2}), we call
Zhang--Yeung's inequality \emph{essentially conditional}. 
(The formal definition  of an essentially conditional linear 
information inequality is given in  Section~\ref{section-def-conditional-inequality}.) 
We show that several other inequalities are also essentially conditional
in the same sense,
\ie   cannot be deduced directly from any unconditional linear inequality.
Besides Zhang--Yeung's  inequality  discussed
above, we prove this property for a conditional inequality  from \cite{Matus99} 
and for three  conditional inequalities  implicitly proven  in \cite{Matus-inf}. 
We also prove one new conditional inequality and show that it is
essentially conditional.

We notice that essentially conditional information inequalities can be divided into two
classes: some of them hold only for \emph{entropic 
points} (\ie points representing  entropies of some distributions), 
other hold also for \emph{almost entropic points} (limits of entropic points). 
Conditional inequalities of 
the second  type (those which hold for almost entropic points) 
have an intuitive geometric meaning.  
We show that the very fact that such inequalities exist,  
implies the seminal theorem of Mat\'{u}\v{s}':
the cone of linear information inequalities is not polyhedral. 
The  physical meaning of  the inequalities of the first type 
(that hold for all entropic but not for all almost entropic points)
remains obscure.

Linear information inequalities can also be studied in the framework 
of Kolmogorov complexity. For unconditional linear information 
inequalities the parallel between Shannon's and Kolmogorov's settings
is very prominent.  It is known that 
the class of unconditional linear inequalities for Shannon entropy
coincides with the class of unconditional linear inequalities for Kolmogorov complexity
(that hold up to an additive logarithmic term), \cite{hrsv}.  
Conditional information inequalities also can be considered in
the setting of Kolmogorov complexity, though
the parallelism between Shannon's and Kolmogorov's conditional 
inequalities is more complicated.  We show that  three essentially conditional
information inequalities are valid, in some natural sense, for Kolmogorov complexity 
(these are the same three conditional inequalities proven to be valid 
for almost entropic points), while  two other conditional  inequalities (which hold for entropic but
not for almost entropic points) do not hold for Kolmogorov complexity. For
another essentially conditional inequality the question remains open, we do not
know whether it holds for almost entropic points and/or for Kolmogorov complexity.


\smallskip

\begin{remark}
It seems that essentially conditional information inequalities are not so far a common 
subject to study. Nevertheless, conditional inequalities are  widely used
in information theory as a helpful tool. 
For instance, conditional information inequalities are involved in 
proofs of conditional independence inference rules, see
\cite{Matus-I,Matus-II,StudenyThesis}.  Also, lower bounds for the
information ratio in secret sharing schemes are based on conditional
information inequalities, \cite{secret-sharing,burton, beimel-orlov}. 
However, in most applications (e.g., in all applications in secret sharing
known to the authors) the used conditional inequalities are just straightforward
corollaries of unconditional  inequalities, like in Examples~\ref{ex:1}--\ref{ex:3} above.
So, in most papers the attention is  not focused on the fact that a \emph{conditional}
inequality for entropies is employed in this or that proof.  
\end{remark} 

\subsection{Organization of the paper}

In this section we briefly formulate our main results and explain the structure of the paper. 
Our study can be roughly divided into $5$ parts corresponding to the following $5$ issues:
\begin{enumerate}
\item formal definitions of conditional and essentially conditional linear inequalities; 
\item the known techniques  and the known  conditional linear information inequalities; 
\item the main result of the paper:  some conditional information inequalities are indeed essentially conditional;
\item essentially conditional inequalities and non-polyhedrality of the cones of almost entropic points;
\item a transposition of conditional information inequalities to algorithmic information theory.
\end{enumerate}
Let us comment this plan in some detail.

\emph{1. Essentially conditional inequalities.}
In Section~\ref{section-preliminaries} we introduce the central definitions of this 
paper~-- we formally define the notions of \emph{conditional} and  \emph{essentially conditional} linear information inequality. We discuss the intuition behind them and the geometric meaning of different types of conditional inequalities.

\emph{2. Non trivial proofs of conditional inequalities.} In Section~\ref{cond-ineq} we prove several nontrivial conditional inequalities (Theorem~\ref{th1} and Theorem~\ref{th-matus-07}).

In Examples~\ref{ex:1}--\ref{ex:3} discussed in the introduction, we presented a trivial way to deduce a conditional 
information inequality from the corresponding unconditional one.
But some known proofs of conditional inequalities do not follow this naive
scheme.
In Section~\ref{cond-ineq} we systematize the known techniques of proofs 
of  conditional linear information inequalities.

Basically, two different techniques are known.  The first technique was
proposed  by Z.~Zhang and R.W.~Yeung in \cite{ZY97},  where they came up
with the first nontrivial conditional inequality.  Later 
the same technique was employed by F.~Mat\'{u}\v{s}  in \cite{Matus99} 
to prove another conditional inequality. In this paper we  use 
a similar argument to prove one more inequality.
Each of these three inequalities involves  four random variables. 
We denote these inequalities~($\I$), ($\II$), and ($\III$) respectively;
the validity  of these inequalities constitute Theorem~\ref{th1} in Section~\ref{cond-ineq}. 
In this paper we prove only  the new Inequality ($\III$) and refer 
the reader  for the proofs of ($\I$) and ($\II$)   to \cite{ZY97} and \cite{Matus99}
respectively.

Three other conditional inequalities are proven implicitly by Mat\'{u}\v{s}
in \cite{Matus-inf}. A different technique is used in this proof;
it is based on  approximation of a conditional inequality with some infinite series 
of unconditional ones\footnote{The unconditional inequalities involved in the proof
are actually  derived themselves
by Zhang--Yeung's method, so the first technique is also implicitly used here.}.
We  formulate this result of Mat\'{u}\v{s}'  in Theorem~\ref{th-matus-07}
(and give an explicit proof). 
In what follows we denote these inequalities~($\IV$--$\VI$).
Each of these three  inequalities involves five random variables. Identifying some
variables in these inequalities, we obtain as a corollary two
nontrivial conditional inequalities ($\IV'$--$\V'$) with four variables which are
of independent interest (see below).  
The main advantage of the technique in 
Theorem~\ref{th-matus-07} is that ($\IV$--$\VI$) and ($\IV'$--$\V'$)
can be proven not only for entropic  but also  for almost entropic points. 

\emph{2. Essentially conditional inequalities.}
In Section~\ref{essentially-cond}  we prove  the main technical result of the paper: we show that several inequalities are essentially conditional (Theorem~\ref{th-not-unconditional}).  

As we mentioned above, Inequalities~($\IV$--$\VI$) hold  for almost entropic points. This is not the case for ($\I$)
and ($\III$). In Theorem~\ref{th-counterexample-aep} of
Section~\ref{sub:section-aepoint} we prove that ($\I$) and
($\III$) are valid only for entropic but do not for almost entropic points.
We actually prove that there exist almost entropic
points that satisfy all unconditional linear inequalities for entropy
but do not satisfy conditional inequalities ($\I$) and $(\III)$.
For ($\II$) the question remains open: we do not know whether it is true for
almost entropic points.

\emph{3. Geometric meaning of essentially conditional inequalities.}
In Section~\ref{sub:section-inf-inequalities} 
we discuss in more technical detail the geometric meaning of inequalities
that hold for almost entropic points. We explain that each essentially conditional inequality that holds for almost entropic points implies the Mat\'{u}\v{s} theorem (Theorem~\ref{th-non-polyhedral-general}): the cone of almost entropic points and  the cone of all (unconditional) linear inequalities 
for entropies for $4$ random variables is not polyhedral if $n\ge4$. 
Basically, our proof is a more geometric explanation of the ideas that appear 
in the original argument by Mat\'{u}\v{s} in~\cite{Matus-inf}.

\emph{4. Conditional inequalities for Kolmogorov complexity.}
In Section~\ref{section-kolm} we discuss essentially conditional inequalities
for Kolmogorov complexity.  In Theorem~\ref{th-matus-07-kolm} we prove that
some counterparts of ($\IV$--$\VI$) are valid for Kolmogorov complexity; in
Theorem~\ref{th-counterexample-kolm} we show that natural  counterparts of
($\I$) and ($\III$) do not hold for Kolmogorov complexity.

\smallskip

Proofs of some technical lemmas are given in Appendix.

\section{Preliminaries}\label{section-preliminaries}

\subsection{Basic definitions and notation\label{sub:section-notation}}

Here we define some notation  (mostly standard)
used in the paper.

\paragraph{Shannon entropy}
All random variables in this paper are discrete random variables with a finite range.
Shannon entropy of a random variable $X$ is defined as
$$\h(X) = \sum\limits_{x}{p(x)\log\frac1{p(x)}}$$
where $p(x) = \prob[X = x]$ for each $x$ in the range of $X$.

\paragraph{Entropy profile}
Let $\mathcal{X} =\{ X_1,\ldots,X_n\}$ be $n$ jointly distributed random
variables. Then each non-empty tuple $X \subseteq \mathcal{X}$ is itself a random variable 
to which can be associated Shannon entropy $\h(X)$. The \emph{entropy profile} of $\mathcal{X}$  
is defined as the vector
$$\vec{H}(\mathcal{X}) = (\h(X))_{\emptyset\neq X \subseteq \mathcal{X}}$$
of entropies for each non-empty subset of random variables, in the lexicographic order.

\paragraph{Entropic points}
A point in $\mathbb{R}^{2^n-1}$ is called \emph{entropic}
if it is the entropy profile of some distribution $\mathcal{X}$. A point is called
\emph{almost entropic} if it belongs to the closure of the set of entropic points.

\paragraph{Linear information inequality}\label{unconditional-inequality}
A linear information inequality for $n$-tuples of random variables
is a linear form with $2^n-1$ real coefficients
$(\kappa_X)_{\emptyset\neq X\subseteq\mathcal {X}}$  
such that  for all jointly distributed random variables $\mathcal{X}=\{X_1,\ldots,X_n\}$
$$\sum\limits_{\emptyset\neq X\subseteq \mathcal{X} }{\kappa_X \h(X)} \ge 0.$$

\paragraph{Box notation}
To make formulae more compact,
we shall use the notation from \cite{Matus-inf}:
$$\square_{AB,CD} := I(C\lon D|A) + I(C\lon D|B) +  I(A\lon B) -  I(C\lon D).$$


\subsection{Conditional information inequalities (the main definitions)}\label{section-def-conditional-inequality}

Linear information inequalities (see \emph{Basic definitions~(d)} above) are the most general properties of Shannon entropy, \ie the inequalities that are valid for all distributions, without any constraints or conditions. 
Our work focuses on the somewhat subtler notion of \emph{conditional} information inequalities. In the introduction, we have already presented several motivating examples   (Examples~1--3 and Inequality (\ref{zy97})). Let us now give a more formal definition of linear conditional information inequalities.
\begin{definition}
Let~$\alpha(\mathcal{X})$~and~$\beta_1(\mathcal{X}), \ldots, \beta_m(\mathcal{X})$ 
be linear functions on entropies  of $\mathcal{X}=(X_1,\ldots,X_n)$ 
$$
\begin{array}{rcl}
\alpha(\mathcal{X})&=&\sum\limits_{\emptyset\not=X\subseteq \mathcal{X}}{\alpha_X\h(X)},\\
\beta_i(\mathcal{X}) &=&\sum\limits_{\emptyset\not=X\subseteq \mathcal{X}}{\beta_{i,X}\h(X)},\ i=1\ldots m
\end{array}
$$
such that  the implication
$$(\beta_i(\mathcal{X}) = 0 \mbox{ for all }i=1,\ldots,m) \Ra \alpha(\mathcal{X}) \ge 0$$
holds for all distributions $\mathcal{X}$. We call this implication  
\emph{a conditional linear information inequality}.
\end{definition}

Examples~\ref{ex:1}--\ref{ex:3} and~(\ref{zy97}) are instances of conditional 
information inequalities.  All these inequalities look quite similar to each other. But 
we claim that  Inequality~(\ref{zy97}) is sharply different from the
inequalities discussed in Examples~1--3. The peculiarity  of (\ref{zy97}) is
that it cannot be deduced directly from any unconditional inequality. In
general, we say that an inequality with linear constraints is \emph{essentially
conditional} if it cannot be extended to any unconditional inequality where
conditions are added with ``Lagrange multipliers''. Let us define this class of
inequalities more precisely.
\begin{definition}
\label{def-essentially-cond}
Let~$\alpha(\mathcal{X})$~and~$\beta_1(\mathcal{X}), \ldots, \beta_m(\mathcal{X})$ 
be linear functions on entropies  of $\mathcal{X}=(X_1,\ldots,X_n)$, and 
$$(\beta_i(\mathcal{X}) = 0 \mbox{ for all }i=1,\ldots,m) \Ra \alpha(\mathcal{X}) \ge 0$$
be a conditional information inequality. We call this implication  
\emph{an essentially conditional linear information inequality}, if
for all $(\lambda_i)_{1\le i\le m}$ the inequality
\begin{equation}\label{eq:def2}
\alpha(\mathcal{X}) + 
\sum\limits_{i=1}^m{\lambda_i \beta_i(\mathcal{X})} \ge 0 
\end{equation}
does not hold (there exists some distribution $\mathcal{X}$ such that~(\ref{eq:def2}) is not satisfied).
\end{definition}

The very fact that some conditional information inequalities are \emph{essentially conditional} is not obvious.
We postpone for a while the proof of existence of such inequalities.
In Section~\ref{essentially-cond}
we will show that Inequality~(\ref{zy97}) as well as several other conditional inequalities are indeed essentially conditional.


\subsection{Geometric interpretation of essentially conditional inequalities}\label{subsection-geometric-meaning}

In what follows we discuss the geometric meaning of linear conditional inequalities of different types. We do not prove here any precise mathematical statements, so the reader can safely skip this section. But we believe that some informal introduction can help grasp the intuition behind conditional inequalities and better understand the technical results in the next sections.

It is known that for every integer $n\ge 1$ the class of all points in $\mathbb{R}^{2^n-1}$ that satisfy all linear information inequalities is exactly the class of all almost entropic points, see \cite{yeung-book}. The class of all almost entropic points is a convex cone, and the (unconditional) linear information inequalities correspond to supporting half-spaces for this cone.

What is the relation between the cone of almost entropic points and the  conditional information inequalities?
It turns out that conditional and essentially conditional inequalities have a very clear geometric interpretation in terms of this cone. However, there is one technical difficulty: all nontrivial conditional inequalities in this text live in a high-dimensional space. Indeed, the simplest example of an essentially conditional inequality involves entropies of $4$ random variables; so, it is an inequality in $\mathbb{R}^{15}$.  We do not provide illustrative pictures for these particular inequalities, however
%
%
to explain the geometric intuition behind conditional information inequalities we simplify the situation. First, we illustrate the meaning of conditional inequalities on $2$-dimensional pictures instead of $15$-dimensional spaces. Second, we draw pictures not with a convex cone but just with some convex geometrical set in the plane, e.g., with a convex polygon. 
Thus, the pictures below are not direct representations of information inequalities. Keep in mind that these pictures must be understood only as metaphorical illustrations.

First of all, we illustrate the geometric meaning of a non-essentially conditional inequality. Assume that
a conditional inequality 
$$(\beta_i(\mathcal{X}) = 0 \mbox{ for all }i=1,\ldots,m) \Ra \alpha(\mathcal{X}) \ge 0$$
is a corollary of some unconditional Inequality~(\ref{eq:def2}).
Typically the \emph{conditions} (linear constraints) $\beta_i(\mathcal{X}) = 0$
correspond to some facet of the cone of almost entropic points. Thus, the conditional inequality corresponds to a (linear) boundary inside this facet.
Such a boundary in a facet of the cone
results from ``traces'' of other unconditional information inequalities (other supporting hyperplanes of the cone).
Geometrically this means that the given conditional inequality  can be extended 
to a hyperplane which makes a boundary to the entire cone.

\begin{figure}[H]		\label{fig-polyg1}
\centering
		\includegraphics[scale=0.8]{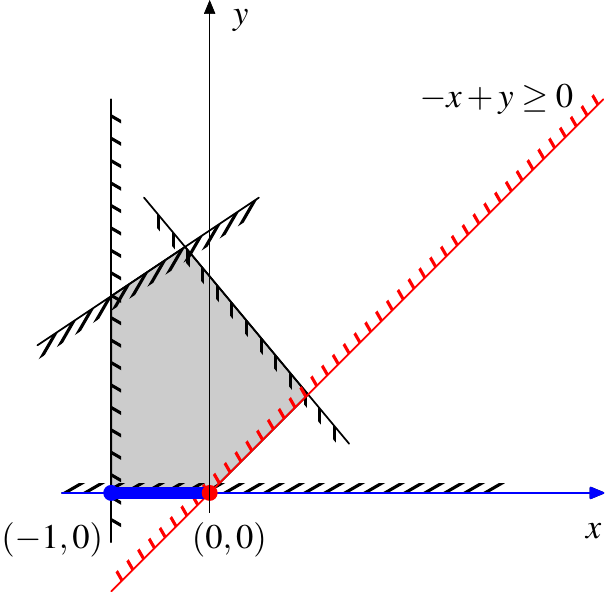}
		\caption{If $y=0$ then $x\le 0$. This conditional inequality follows from $-x+y\ge0$.}		

\end{figure}

Let us illustrate this effect on a simple $2$-dimensional picture. In Fig.~1 we have a closed convex polygon
in the plane, whose boundary is made of $5$ lines. In other words,  this polygon can be represented as an intersection of $5$  half-spaces ($5$ unconditional linear inequalities).
Now we restrict our attention to the line $y=0$ and see that a point with coordinates $(x,0)$ belong to the polygon, if and only if $-1\le x\le 0$. In particular, we have a conditional inequality for this polygon:
\begin{equation}
\label{eq:cond-ineq-picture}
\mbox{if }y=0\mbox{ then }x\le 0. 
\end{equation}
Where does the bound $x\le 0$ come from? The answer is obvious: the point 
$(0,0)$ is the ``trace'' in the line $y=0$ obtained from the unconditional inequality $-x+y\ge 0$ . That is,
conditional Inequality~(\ref{eq:cond-ineq-picture})
can be extended to the unconditional inequality $-x+y\ge0$.

Let us consider a more complex example, see Fig.~2. In this picture the geometrical figure (a closed convex area shaded grey) is not a polygon, its boundary contains a curve arc. We assume that the tangent line to this curve at the point $(0,0)$ is horizontal.

\begin{figure}[H] \label{fig-polyg2}
\centering
		\includegraphics[scale=0.8]{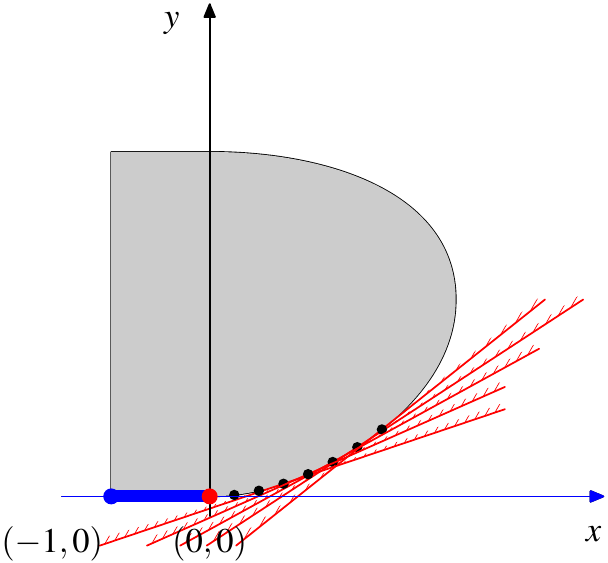}
		\caption{If $y=0$ then $x\le 0$. This conditional inequality is implied by an \emph{infinite} family of tangent half-planes.}	
\end{figure}

Similarly to the previous example, 
we restrict our attention to the line $y=0$;  we see that a point with coordinates $(x,0)$ belongs to the gray figure, if and only if $-1\le x\le 0$. That is, we have again a conditional Inequality~(\ref{eq:cond-ineq-picture}).
But in this case the conditional inequality cannot be extended to any unconditional linear inequality (for every $\lambda\ge 0$ the  inequality $-x+\lambda y\ge 0$ is not valid for the grey area in the picture). Thus, here we deal with an essentially conditional inequality.

In some sense, this inequality follows from an \emph{infinite} family of unconditional inequalities: (\ref{eq:cond-ineq-picture})  is implied by the family of all supporting half-place that are tangent to the grey area in the picture.
This phenomenon can occur for a convex closed body only if this body is not polyhedral (not polygonal in the $2$-dimensional case).

Next, we consider another important example. In Fig.~3 the grey area is again a convex set, but it is not closed. 
We assume that the part of its boundary shown by the bold lines belongs to the figure while
the part of the boundary shown by  the dashed lines does not. 
Then for the grey area on the picture we have again the same conditional Inequality~(\ref{eq:cond-ineq-picture}).
Obviously, this inequality is essentially conditional: it cannot be extended to an inequality 
$-x+\lambda y\ge 0$ with any $\lambda$. Moreover,~(\ref{eq:cond-ineq-picture}) holds for the grey area but does not hold for its closure, e.g.,  the point $(1,0)$
does not satisfy this conditional inequality.
\begin{figure}[H] \label{polyg1}
\centering
		\includegraphics[scale=0.8]{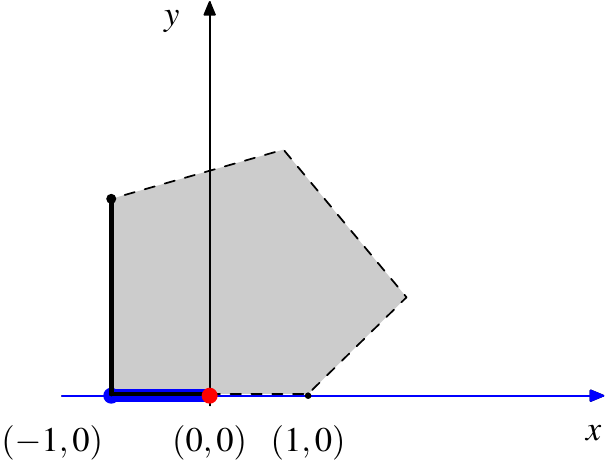}
		\caption{If $y=0$ then $x\le 0$. This conditional inequality does not hold for the \emph{closure} of the grey area.}		

\end{figure}

The three simple examples above are illustrative and helpful to understand 
the geometry of conditional linear information inequalities. Each conditional linear inequality for entropies is a linear inequality that is valid on some ``facet'' on the cone of all entropic points (the co-dimension of this facet corresponds to the number of linear constraints in this conditional inequality). 
\begin{itemize}
\item If an inequality is not essentially conditional, then it can be extended to an unconditional inequality which corresponds to a  supporting half-space for the entire cone of all almost entropic points, similarly to the example in Fig.~1.
\item If this inequality is essentially conditional, then two different cases are possible:
\begin{enumerate}[(a)]
\item If a conditional inequality is valid not only for entropic but also for almost entropic points (for the closure of the set of entropic points), then this inequality follows from an infinite family of unconditional inequalities (an infinite family of supporting half-spaces that are tangent to the cone of almost entropic points), similarly to the example in Fig.~2. In the next sections we  discuss several examples of essentially conditional information inequalities of this type. From the fact that such conditional information inequalities exist it follows that the cone of almost entropic points (for $n\ge 4$ random variables) is not polyhedral (\ie every essentially conditional inequality of this type provides an alternative  proof of the Mat\'{u}\v{s} theorem from \cite{Matus-inf}). In Section~\ref{sub:section-inf-inequalities} we discuss this result in more detail.

Note that the geometry of the cone of almost entropic points is not completely understood yet. For instance, it remains unknown whether this cone has countably or uncountably many tangent half-planes; in the first case the surface of the cone consists of countably many flat facets; in the second one, some parts of the surface are curved and then some non linear  inequalities should play the key role. We believe that studying of essentially conditional inequalities can shade some light on this problem, though for the time being we cannot make any  specific conjecture. 

\item If a conditional inequality is not valid for (some) almost entropic points,  then its geometry is somehow similar to  the example in Fig.~3.  The interpretation of this type of essentially conditional information inequalities still remains unclear. A similar phenomenon was studied in \cite{Matus-piecewise}: there exist  nonlinear (piecewise linear) conditional information inequalities that are valid for all entropic but not for all almost entropic points.
\end{enumerate}
\end{itemize}
The zoo of all known essentially conditional inequalities is not very large so far:
\begin{itemize}
\item Zhang-Yeung's inequality from \cite{ZY97} and the conditional inequality originally proven in \cite{condineq} (see in the next section inequalities ($\I$) and~($\III$) respectively): these essentially conditional inequalities are valid for all entropic but not for all almost entropic points, \emph{cf.} Fig.~3.
\item Mat\'{u}\v{s}' inequalities implicitly proven in  \cite{Matus-inf} (inequalities ($\IV$-$\VI$) and their corollaries ($\IV'$-$\V'$) in the next section): essentially conditional inequalities that are valid  for all almost entropic points, \emph{cf.} Fig.~2. These inequalities reflect the fact 
that the cone of almost entropic points for $n\ge 4$ random variables is not polyhedral.
\item Another inequality proven by Mat\'{u}\v{s} in \cite{Matus99} (Inequality ($\II$) in the next section): the question whether this inequality is valid for almost entropic points remains open. 
\end{itemize}

Thus, we  have defined  the notion of an essentially conditional linear information inequality, and understood the basic geometry behind such inequalities. In the next section we discuss in more detail specific examples of conditional information inequalities. 

\section{Nontrivial Proofs of Conditional  Inequalities}\label{cond-ineq}

As we mentioned in the previous section, several  known conditional information inequalities were proven in a nontrivial way (they were not deduced directly from any unconditional inequality). 
Two different techniques are used to prove such inequalities: a direct application of Zhang--Yeung's method 
and the method of approximation by unconditional inequalities, proposed by Mat\'{u}\v{s}. In this section we systemize the known results and put them in two theorems.

\subsection{The first approach: Zhang--Yeung's method}

Most known proofs of non-Shannon-type information inequalities are based on the
method proposed by Z.~Zhang and R.W.~Yeung in \cite{ZY97} and \cite{ZY98},
see a general exposition  in  \cite{yeung-book,copy-trick}.  
The basic ideas of this method can be explained as follows.  Let us have $n$ jointly
distributed random variables.  We take the marginal distributions for the given
random variables and use them to assemble one or two new distributions. The new
distributions may involve more than $n$ random variables  (we can take several
``copies'' of the same components from the original distribution). Then we
apply some known linear information inequality to the constructed distributions, or use the non-negativity of the Kullback\textendash{}Leibler divergence. 
We express the resulting inequality in terms of entropies of the original
distribution, and get a new inequality.

This type of argument was used in  \cite{ZY97,Matus99,condineq} to prove
some conditional information inequality. We unite these results in the following theorem.
\begin{theorem}\label{th1}
For every distribution $(A,B,C,D)$ 
\begin{itemize}
\item[($\I$)]  if $I(A\lon B|C) = I(A\lon B)=0$, then 
  $\square_{AB,CD}\ge0$,

\smallskip

\item[($\II$)] if $I(A\lon B|C) = I(B\lon D|C)=0$,  then
    $\square_{AB,CD}\ge0$,

\smallskip

\item[($\III$)] if $I(A\lon B|C) = \h(C|A,B)=0$, then 
    $\square_{AB,CD}\ge0$. 
\end{itemize}
\end{theorem}
\begin{remark}  
Note that the linear inequality in statement~($\I$) can be rewritten as $I(C\lon D)\le  I(C\lon D|A) + I(C\lon D|B)$ 
since $\square_{AB,CD} = I(C\lon D|A) + I(C\lon D|B)-I(C\lon D)$ under the restriction $I(A\lon B)=0$.  
Thus,  ($\I$) is just a new name for Inequality~(\ref{zy97})  discussed in Introduction.
\end{remark}
\emph{Historical remarks:} Inequality~($\I$) was
proven by Z.~Zhang and R.W.~Yeung in \cite{ZY97}, ($\II$) was proven by
F.~Mat\'{u}\v{s} in \cite{Matus99}, and ($\III$) was originally obtained in
\cite{condineq}.  In this paper we prove only~($\III$). 

\begin{IEEEproof}[Proof of $(\III)$]
The argument consists of two steps: `enforcing conditional independence' and
`elimination of conditional entropy'.  Let  $(A,B,C,D)$ be jointly distributed
random variables. The first trick of the argument is a suitable
transformation of this  distribution. We keep the same distribution on the
triples $(A,C,D)$ and $(B,C,D)$ but make $A$ and $B$ independent conditional on
$(C,D)$. Intuitively it means that we first choose at random (using the old
distribution) values of $C$ and $D$; then given fixed values of $C,D$ we
independently choose at random $A$ and $B$ (the conditional distributions of
$A$ given $(C,D)$ and $B$ given $(C,D)$ are the same as in the original
distribution). More formally, if $p(a,b,c,d)$ is
the original distribution (we denote $p(a,b,c,d)=\prob[A=a,B=b,C=c,D=d]$), then the new distribution $p'$ is defined as
\begin{IEEEeqnarray*}{rcl}
p'(a,b,c,d)&=&
 \frac{p(a,c,d)\cdot p(b,c,d)}{p(c,d)}
\end{IEEEeqnarray*}
(for all values $(a,b,c,d)$ of the four random variables). 
With some abuse of notation we denote the new random variables 
by $A',B',C,D$. From the construction ($A'$ and $B'$ are independent given $C,D$) it follows that
 $$
 \h(A',B',C,D) = \h(C,D) + \h(A'|C,D) + \h(B'|C,D)
 $$
Since $(A',C,D)$ and $(B',C,D)$ have exactly the same distributions as the original 
$(A,C,D)$ and $(B,C,D)$ respectively, we have
 $$
 \h(A',B',C,D) = \h(C,D) + \h(A|C,D) + \h(B|C,D)
 $$
The same entropy can be bounded in another way using the chain rule and since conditioning  decreases entropy:
$$
 \h(A',B',C,D) \le \h(D) + \h(A'|D)
+ \h(B'|D)+ \h(C|A',B')
$$
Notice that the conditional entropies $\h(A'|D)$ and $\h(B'|D)$ are equal to $\h(A|D)$ and $\h(B|D)$
respectively (we again use the fact that $A',D$ and $B',D$ have the same distributions as $A,D$
and $B,D$ in the original distribution). Thus, we get
\begin{eqnarray*}
\h(C,D) + \h(A|C,D) + \h(B|C,D) &\le&     
  \h(D) + \h(A|D)+ \h(B|D)+ \h(C|A',B')
\end{eqnarray*}
It remains to estimate the value $\h(C|A',B')$. We will show that it is zero (and this is the
second trick used in the argument).

Here we will use the two conditions of the theorem. We say that some values $a,c$ 
($b,c$ or $a,b$ respectively)
are \emph{compatible} if in the original distribution these values can appear together, \ie
$p(a,c)>0$ ($p(b,c)>0$ or $p(a,b)>0$ respectively). 
Note that in general for some distributions $(A,B,C,D)$ it can happen that some value of $A$ and some value of $B$ are each compatible with one and the same value of $C$, but not compatible with each other. However, such a phenomenon is impossible for our distribution. Indeed, we are given that $A$ and $B$ are independent given $C$. This property implies that if some values $a$ and $b$ are compatible with the same value $c$, then these $a$ and $b$ are also compatible with each other.

In the new distribution $(A',B',C,D)$, values of $A'$ and $B'$ are compatible with each other \emph{only if}
they are compatible with some value of $C$; hence, these values must also be compatible with each other for the original
distribution $(A,B)$. Further, since $\h(C|A,B)=0$, for each pair of compatible values of $A,B$ there exists only one
value of $C$. Thus, for a random pair of values of $(A',B')$ with probability one there exists only one value of $C$.
In a word, in the new distribution $\h(C|A',B')=0$.  

Summarizing our arguments, we get 
$$
\h(C,D) + \h(A|C,D) + \h(B|C,D) 
 \le \h(D) + \h(A|D)+ \h(B|D),
$$
which is equivalent to 
\begin{IEEEeqnarray*}{rCl+x*}
 I(C\lon D)&\le& I(C\lon D|A) + I(C\lon D|B) + I(A\lon B) &\IEEEQED
\end{IEEEeqnarray*}\let\IEEEQED\relax
\end{IEEEproof}
\vspace{-0.5cm}

\subsection{The second approach: a conditional inequality as a limit of unconditional ones}
\label{sub:section-limit-of-uncond}

Another group of conditional inequalities was proven implicitly
in~\cite{Matus-inf}. In this argument a conditional inequality is obtained as 
a limit of some infinite family of unconditional inequalities.
\begin{theorem}[F.~Mat\'{u}\v{s}]\label{th-matus-07}
For every distribution $(A,B,C,D,E)$
\indent
\begin{itemize}
\item[$(\IV)$] if $ I(A\lon D|C)=I(A\lon C|D)=0$,  then   
$ \square_{AB,CD} + I(A\lon C|E)+I(A\lon E|C)\ge0,$

\smallskip

\item[$(\V)$] if $I(B\lon C|D)=I(C\lon D|B)=0$,  then  
 $ \square_{AB,CD} + I(B\lon C|E)+I(C\lon E|B)\ge0,$

\smallskip

\item[$(\VI)$] if $I(B\lon C|D)=I(C\lon D|B)=0$, then 
$ \square_{AB,CD} + I(C\lon D|E)+I(C\lon E|D)\ge0$.
\end{itemize}
These inequalities hold not only for entropic but also for almost entropic points.
\end{theorem}
\begin{IEEEproof}
The following series of unconditional inequalities were proven in~\cite{Matus-inf} for all $k=1,2,\dotsc$
\begin{IEEEeqnarray*}{ll}\label{uncond-ineq}
(i)&  \square_{AB,CD} + I(A\lon C|E)+I(A\lon E|C) +
 \frac{1}{k}I(C\lon E|A)
 +\frac{k-1}{2}(I(A\lon D|C)+I(A\lon C|D)) \ge0, 
 \\
(ii)& \square_{AB,CD} + I(B\lon C|E)+I(C\lon E|B) +
\frac{1}{k}I(B\lon E|C) 
 + \frac{k-1}{2}(I(B\lon C|D)+I(C\lon D|B))\ge0, 
 \\
(iii)\ \ &  \square_{AB,CD} + I(C\lon D|E)+I(C\lon E|D) +
\frac{1}{k}I(D\lon E|C) 
+\frac{k-1}{2}(I(B\lon C|D)+I(C\lon D|B))\ge0 
\end{IEEEeqnarray*}
(inequalities ($i$), ($ii$), and ($iii$) correspond to the three claims of Theorem~2 in~\cite{Matus-inf}, for
$\reflectbox{\rotatebox[origin=c]{90}{$\boxminus$}}$ equal to
$\square_{14,23}$, $\square_{13,24}$, and $\square_{12,34}$ respectively).

The constraints in ($\IV$), ($\V$) and ($\VI$) imply that the terms with the
coefficient $ \frac{k-1}{2}$ in ($i$), ($ii$) and ($iii$), respectively, are
equal to zero. The terms with the coefficient $\frac{1}{k}$ vanish as $k$
tends to infinity.  So in the limit we obtain from ($i$--$iii$) the required
inequalities ($\IV$--$\VI$).

Note that linear inequalities ($i$), ($ii$), ($iii$) hold for all points in the
cone of almost entropic points. Hence, the limits ($\IV$--$\VI$) are also valid
for almost entropic points.

\end{IEEEproof}\label{th-cond-approx}
We can restrict this theorem to $4$-variables distributions and get simpler and
more symmetric inequalities:
\begin{corollary}  \label{corollary-matus-ineq}
For every distribution $(A,B,C,D)$  
\begin{IEEEeqnarray*}{rlClcl}
(\IV')\  &  I(A\lon D|C)=I(A\lon C|D)=0 &\Ra& \square_{AB,CD}\ge0,\\
(\V')\  & I(B\lon C|D)=I(C\lon D|B)=0& \Ra& \square_{AB,CD}\ge0.
\end{IEEEeqnarray*}
These conditional inequalities  are also valid for almost entropic points.
\end{corollary}
\begin{IEEEproof}
Inequality ($\IV'$) is a  restriction of ($\IV$)  for the cases when $e=d$.
Inequality ($\V'$) can be obtained as a restriction of  ($\V$)  or ($\VI'$) for $E=D$ and
$E=B$ respectively.  
\end{IEEEproof}
\subsection{A direct proof of ($\IV'$)}

In this section we show that conditional Inequality ($\IV'$) from Corollary~\ref{corollary-matus-ineq}
can be proven by a  simpler direct argument. In what follows  we  deduce ($\IV'$) (for entropic points) 
from two elementary lemmas.  

\begin{lemma}[Double Markov Property] \label{lemma-iv-1}
Assuming that $$I(X\lon Z|Y)=I(Y\lon Z|X)=0,$$  there exists a random variable $W$ (defined on
the same probabilistic space as $X,Y,Z$) such that
\begin{itemize}
\item $\h(W|X) = 0$,
\item $\h(W|Y) = 0$,
\item $I(Z\lon X,Y| W)=0$.
\end{itemize}
\end{lemma}
This lemma is an exercise in~\cite{csiszar-korner}; see also Lemma~4 in \cite{MMRV}.
For completeness we prove this lemma in Appendix.

\begin{remark} If $I(X\lon Z|Y)=I(Y\lon Z|X)=0$ and $I(Z \lon X,Y| W)=0$, then 
$I(X\lon Y | W)=I(X\lon Y|W,Z)$. Thus, we get from Lemma~\ref{lemma-iv-1}
a random variable $W$ that functionally depends on $X$ and on $Y$,
and $I(X\lon Y | W)=I(X\lon Y|W,Z)$.
\end{remark}
\begin{lemma}\label{lemma-iv-2}
Assume there exists a random variable $W$ such that
\begin{itemize}
\item $\h(W | X) = 0$,
\item $\h(W | Y) = 0$,
\item $I(X\lon Y | W)=I(X\lon Y|W,Z)$.
\end{itemize}
Then for any random variable $V$, the Ingleton inequality $\square_{VZ,XY}\ge0$ holds.
\end{lemma}
\begin{IEEEproof}[Proof of lemma~\ref{lemma-iv-2}]
Since $\h(W | X) = \h(W|Y)=0$,
we get
$$
I(X\lon Y) = \h(W)+ I(X\lon Y|W),
$$
where the value $I(X\lon Y|W)$ can be substituted by $I(X\lon Y|W,Z)$.
Further, for any triple of random variables $V,W,Z$ we have
$$
\h(W) \le \h(W|V)+ \h(W|Z)+I(Z\lon V).
$$
This inequality  follows from the equality  $\h(W) = \h(W|V) + \h(W|Z) + I(Z\lon V) - I(Z\lon V|W) - \h(W|Z,V)$
(see also \cite[proof of inequality~(6)]{hrsv}).
Then, we use again the assumption $\h(W | X) =\h(W|Y)= 0$ and get
$$
\h(W|Z) + I(X\lon Y|W,Z)  = I(X\lon Y|Z)
$$
and
$$
\h(W|V) \le I(X\lon Y|V).
$$
The sum of these relations results in $\square_{VZ,XY}\ge0$.
\end{IEEEproof}
Inequality ($\IV'$) follows immediately from these two lemmas (substitute $V,Z,X,Y$ with $A,B,C,D$ respectively). 

\begin{remark}
This argument employs only Shannon-type inequalities. The single step which
cannot be reduced to a combination of basic inequalities is the lemma about
Double Markov property: the main trick of this proof is the construction of
a new random variable $W$ in Lemma~\ref{lemma-iv-1}.
\end{remark}
\begin{remark}
The proof presented in this section is simpler, but 
the more involved argument  explained in Section~\ref{sub:section-limit-of-uncond}
gives much more information about ($\IV'$). Indeed,
the argument based on approximation with unconditional linear inequalities
implies that  ($\IV'$) holds not only for entropic but also for almost
entropic points.
\end{remark}

\section{Why Some Inequalities are Essentially Conditional\label{essentially-cond}}

Now we are ready to prove the main result of this paper. We claim that  inequalities ($\I$--$\VI$) and ($\IV'$--$\V'$)  are essentially  conditional (in the sense of  Definition~\ref{def-essentially-cond}, p.~\pageref{def-essentially-cond}).
\begin{theorem} \label{th-not-unconditional}
Inequalities  ($\I$--$\VI$) and ($\IV'$--$\V'$) are \emph{essentially} conditional. 
\end{theorem}

\begin{remark}
From the fact that ($\IV'$--$\V'$) are essentially conditional, it follows
immediately that ($\IV$--$\VI$) are also essentially conditional. Thus, it
remains to prove the theorem for  ($\I$--$\III$) and ($\IV'$--$\V'$).
\end{remark}

\subsection{Proof of Theorem~\ref{th-not-unconditional} based on binary (counter)examples}

\emph{Claim 1.} \label{claim1}
For any reals $\lambda_1,\lambda_2$ the inequality
 \begin{equation}
 \label{ineq-conjecture-mmrv}
\begin{aligned}
 I(C\lon D) \le I(C\lon D|A) + I(C\lon D|B) 
  +\lambda_1 I(A\lon B|C) + \lambda_2 I(A\lon B) 
\end{aligned}
\end{equation}
does not hold for some distributions $(A,B,C,D)$, \ie ($\I$) is essentially conditional.

\begin{IEEEproof}
For all $\varepsilon\in[0,1]$, consider the following  joint distribution of binary variables 
$(A,B,C,D)$:
$$
\begin{array}{rcc}
 \prob[A=0,\ B=0,\ C=0,\  D=1] &=&  (1-\varepsilon)/4, \\
 \prob[A=0,\ B=1,\ C=0,\  D=0] &=& (1-\varepsilon)/4, \\
 \prob[A=1,\ B=0,\ C=0,\  D=1] &=& (1-\varepsilon)/4,  \\
 \prob[A=1,\ B=1,\ C=0,\  D=1] &=&(1-\varepsilon)/4,  \\
 \prob[A=1,\ B=0,\ C=1,\  D=1] &=& \varepsilon. 
\end{array}
$$
For each value of $A$ and for each value of $B$, the value
of at least one of variables $C,D$ is uniquely determined: if $A=0$ then $C=0$;
if $A=1$ then $D=1$; if $B=0$ then $D=1$; and if $B=1$ then $C=0$. Hence,
$I(C\lon D|A)=I(C \lon D|B)=0$.  Also it is easy to see that $I(A\lon B|C)=0$. Thus, if~(\ref{ineq-conjecture-mmrv})
is true, then
$
I(C\lon D) \le \lambda_2 I(A\lon B).
$

Denote the right-hand and left-hand sides of this inequality by $L(\varepsilon)=I(C\lon D)$ and 
$R(\varepsilon)=\lambda_2 I(A\lon B)$. Both  functions $L(\varepsilon)$ and $R(\varepsilon)$
are continuous, and $L(0)=R(0)=0$ (for $\varepsilon = 0$
both sides of the inequality are equal to $0$). However the asymptotics of 
$L(\varepsilon)$ and $R(\varepsilon)$ as $\varepsilon\to 0$, are different:
it is not hard to check that  
$L(\varepsilon) = \Theta(\varepsilon)$, but  $R(\varepsilon) = O(\varepsilon^2)$.
From~(\ref{ineq-conjecture-mmrv}) it follows
$
\Theta(\varepsilon) \le O(\varepsilon^2),
$
which is a contradiction.
\end{IEEEproof}

\emph{Claim 2.} For any reals $\lambda_1,\lambda_2$ the inequality
\begin{eqnarray} \label{false-conj-3}
\begin{aligned}
 I(C\lon D)\le I(C\lon D|A)+I(C\lon D|B) + I(A\lon B)
 + \lambda_1 I(A\lon B|C) + \lambda_2 I(B\lon D|C)
\end{aligned}
\end{eqnarray}
does not hold for some distributions $(A,B,C,D)$, \ie ($\II$) is essentially conditional.

\begin{IEEEproof}
 For the sake of contradiction we consider the following  joint distribution of binary variables 
 $(A,B,C,D)$
for every value of $\varepsilon\in[0,\frac13]$:
$$
\begin{array}{rcc}
 \prob[A=0,\ B=0,\ C=0,\  D=0] &=& 3\varepsilon,\\
 \prob[A=1,\ B=1,\ C=0,\  D=0] &=& 1/3-\varepsilon,\\
 \prob[A=1,\ B=0,\ C=1,\  D=0] &=& 1/3-\varepsilon, \\
 \prob[A=0,\ B=1,\ C=0,\  D=1] &=& 1/3 -\varepsilon.
\end{array}
$$
For any $\varepsilon$, we see that $ I(C\lon D|A) = I(C\lon D|B) = 0$ and $I(A\lon B|C) = I(B\lon D|C)$.

When $\varepsilon = 0$, the first atom of the distribution disappears and  the distribution satisfies $I(C\lon D)_{|\varepsilon=0} = I(A\lon B)_{|\varepsilon=0} = I_0$.
However, the asymptotics of $I(C\lon D)$ and $ I(A\lon B)$ are different. Indeed,  computations provide $I(C\lon D) = I_0 + \Theta(\varepsilon)$ while  $I(A\lon B) =  I_0 + \Theta(\varepsilon\log(\varepsilon))$. Moreover, 
$I(A\lon B|C)= O(\varepsilon)$,
thus when substituting  in~(\ref{false-conj-3}) we obtain
$$
I_0+\Theta(\varepsilon) \le  I_0 + \Theta(\varepsilon\log(\varepsilon) + O((\lambda_1+\lambda_2) \varepsilon),
$$
and get a contradiction as $\varepsilon\to 0$ .
\end{IEEEproof}

\emph{Claim 3.} For any reals $\lambda_1,\lambda_2$ the inequality
\begin{eqnarray}\label{false-conj-2}
\begin{aligned}
I(C\lon D)\le I(C\lon D|A)+I(C\lon D|B) + I(A\lon B)+
\lambda_1 I(A\lon B|C) + \lambda_2 \h(C|A,B)
 \end{aligned}
\end{eqnarray} 
does not hold for some distributions $(A,B,C,D)$, \ie ($\III$) is essentially conditional.

\begin{IEEEproof}
For every value of $\varepsilon\in[0,\frac12]$ we consider 
the following  joint distribution of binary variables $(A,B,C,D)$:
$$
\begin{array}{rcl}
 \prob[A=1,\  B=1,\ C=0,\ D=0] &=& 1/2 -\varepsilon,\\
 \prob[A=0,\  B=1,\ C=1,\ D=0] &=& \varepsilon,\\
 \prob[A=1,\  B=0,\ C=1,\ D=0] &=& \varepsilon, \\
 \prob[A=0,\  B=0,\ C=1,\ D=1] &=& 1/2 -\varepsilon.
\end{array}
$$
First, it is not hard to check that $I(C\lon D|A)=I(C\lon D|B)=\h(C|A,B)=0$
for every $\varepsilon$. Second, 
\begin{IEEEeqnarray*}{rcl}
 I(A\lon B)&=& 1 + (2-2/\ln 2)\varepsilon + 2\varepsilon\log \varepsilon + O(\varepsilon^2),\\
 I(C\lon D)&=& 1 +(4-2/\ln 2)\varepsilon + 2\varepsilon\log \varepsilon  + O(\varepsilon^2),
\end{IEEEeqnarray*}
so $I(A\lon B)$ and $I(C\lon D)$ both tend to $1$ as $\varepsilon\to 0$, but their asymptotics are different. 
Similarly, 
 $$
 I(A\lon B|C) = O(\varepsilon^2).
$$
It follows from~(\ref{false-conj-2}) that
$$
 2\varepsilon +O(\varepsilon^2) \le 
  O(\varepsilon^2) +  O(\lambda_1 \varepsilon^2),
$$
and with any $\lambda_1$ we get a contradiction for small enough $\varepsilon$.
\end{IEEEproof}

\emph{Claim 4.}  For any reals $\lambda_1,\lambda_2$ the inequality
 \begin{equation}\label{eq:claim4}
\square_{AB,CD} +\lambda_1 I(A\lon C|D) + \lambda_2 I(A\lon D|C) \ge 0
\end{equation}
does not hold for some distributions, \ie ($\IV'$) is essentially conditional.
\begin{IEEEproof}
For all $\varepsilon\in[0,\frac14]$, consider the following joint distribution of binary variables 
$(A,B,C,D)$:
$$
\begin{array}{rcl}
 \prob[A=0,\ B=0,\ C=0,\  D=0] &=& \varepsilon, \\
 \prob[A=1,\ B=1,\ C=0,\  D=0] &=& \varepsilon, \\
 \prob[A=0,\ B=1,\ C=1,\  D=0] &=& \frac14,  \\
 \prob[A=1,\ B=1,\ C=1,\  D=0] &=& \frac14-\varepsilon,  \\
 \prob[A=0,\ B=0,\ C=0,\  D=1] &=& \frac14-\varepsilon, \\
 \prob[A=1,\ B=0,\ C=0,\  D=1] &=& \frac14.
\end{array}
$$
For this distribution, we have $$I(A\lon C|D) = I(A\lon D|C) = \Theta(\varepsilon^3),$$ while 
$$
\square_{AB,CD}  =  -\frac{2}{\ln 2}\varepsilon^2+ O(\varepsilon^3).
$$
Thus, Inequality~\eqref{eq:claim4} rewrites to 
\begin{eqnarray*}
\begin{aligned}
\square_{AB,CD} + \lambda_1 I(A\lon C|D) + \lambda_2 I(A\lon D|C) 
 = -\frac{2}{\ln 2}\varepsilon^2+ O((\lambda_1+ \lambda_2)\varepsilon^3) \ge0,
\end{aligned}
\end{eqnarray*}
but the left-hand side is negative for $\varepsilon$ small enough.
\end{IEEEproof}

\emph{Claim 5.}\label{claim5} For any reals $\lambda_1,\lambda_2$ the inequality
 \begin{equation}\label{eq:claim5}
\square_{AB,CD} +\lambda_1 I(B\lon C|D) + \lambda_2 I(C\lon D|B) \ge 0
\end{equation}
does not hold for some distributions, \ie ($\V'$) is essentially conditional.
\begin{IEEEproof}
For all $\varepsilon\in[0,\frac12]$, consider the following joint distribution for 
$(A,B,C,D)$: 
$$
\begin{array}{rcl}
 \prob[A=0,\ B=0,\ C=0,\  D=0] &=& \frac12-\varepsilon, \\
 \prob[A=0,\ B=1,\ C=0,\  D=1] &=& \frac12- \varepsilon, \\
 \prob[A=1,\ B=0,\ C=1,\  D=0] &=& \varepsilon,  \\
 \prob[A=1,\ B=1,\ C=0,\  D=0] &=& \varepsilon.
\end{array}
$$
For this distribution we have $I(C\lon D|A) = I(C\lon D|B) = I(A\lon B) = 0$, and 
$$
I(C\lon D) = \varepsilon + O(\varepsilon^2) \mbox{ and } I(B\lon C|D) = O(\varepsilon^2).
$$
Therefore, the left-hand side of Inequality~\eqref{eq:claim5} rewrites to
\[
\square_{AB,CD} +\lambda_1 I(B\lon C|D) + \lambda_2 I(C\lon D|B)
= -\varepsilon + O(\lambda_1\varepsilon^2),
\]
which is negative for $\varepsilon$ small enough.
\end{IEEEproof}

\subsection{Geometric (counter)example and another proof of Theorem~\ref{th-not-unconditional} for ($\I$) and ($\III$)}
\label{geometric-counterexample}

On the affine plane over a finite field $\mathbb{F}_q$, consider a random quadruple $(A,B,C,D)_q$ of geometric objects defined as follows.

\begin{itemize}

\item First, choose a random (non-vertical) line $C$ defined by the equation
$y=c_0 + c_1x$ (the coefficients $c_0$ and $c_1$ are independent random
elements of the field $\mathbb{F}_q$); 

\item then pick independently and uniformly  two points  
$A$ and $B$ on the line $C$ (technically, $A=(a_x,a_y)\in \mathbb{F}^2_q$ and $B=(b_x,b_y)\in\mathbb{F}^2_q$; since the points on the line are chosen independently, they coincide with each other with probability  $1/q$);

\item pick uniformly at random a parabola $D$ in the set of all
non-degenerate parabolas $y=d_0+d_1x+d_2x^2$ (where
$d_0,d_1,d_2\in\mathbb{F}_q, d_2\neq 0$ are chose uniformly) that intersect $C$ exactly at
points $A$ and $B$ (if $A=B$ then $C$ is the tangent\footnote{Here, the tangent line to $D$ at point $A$ is the unique line that goes through $A$ but no other point of $D$} line to $D$ at $B$).
\end{itemize}
A typical quadruple is shown in Figure~4.
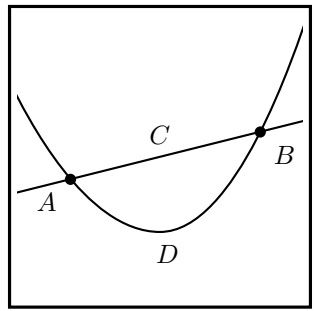
\begin{figure}[H]
\centering
\label{typquad}
\begin{tikzpicture}[scale=1.0]

 \draw[very thick] (0,0) rectangle (4,4);
 \clip (0.1,0.1) rectangle (3.9,3.9);

 \draw[thick] (0,3) parabola bend (2,1) (4,4);
 \draw[thick] (0,1.5) -- (4,2.5);

\node[fill=black,inner sep=1.5pt,label=below left:$A$,circle] 
      (A) at (0.81,1.70) {};
\node[fill=black,inner sep=1.5pt,label=below right:$B$,circle] 
      (B) at (3.333,2.33) {};
\node[above] (C) at (2,2.05) {$C$};
\node[below] (D) at (2.1,0.95) {$D$};
\end{tikzpicture}
\caption{A typical configuration of random objects $(A,B,C,D)$.}
\end{figure}
By  construction, if the line $C$ is known, then
the points $A$ and $B$ are independent, \ie
$$I(A\lon B|C) = 0.$$
Similarly, $A$ and $B$ are independent when $D$ is known, and also
$C$ and $D$ are independent given $A$ and given
$B$ (when an intersection point is given, the line does not give more
information about the parabola), \ie
$$
I(A\lon B| D) = I(C\lon D|A)=I(C\lon D|B)=0.
$$

The mutual information between $C$ and $D$ is approximately $1$
bit because  randomly chosen line and parabola intersect   
iff the discriminant of the corresponding equation is a quadratic residue, 
which happens almost half of the time. We provide next a more accurate computation:

Given $D$, values of $A$ and $B$ are independent.
There are $q^2$ pairs of possible points $A,B$. Each pair is associated with a unique  $C$.
Let us fix some value of parabola $D$. For every points $x$ on the graph of this parabola, 
the probability of the event $A=B=x$ is equal to $1/q^2$.  
If $A$ and $B$ coincides, then $C$ must be a tangent line to $D$.
Hence, for a fixed value of $D$ each of the $q$ possible tangent lines $C$ occurs with probability $1/q^2$.
On the other hand, on the graph of parabola there are $\binom{q}{2}$ pairs of non-coinciding points. 
Each of these pairs is chosen with probability $2/q^2$ (the factor $2$
corresponds to choice between two options: one of these points is chosen as $A$
and another as $B$ or vice versa). Hence, each non-tangent line $C$ is chosen with probability $2/q^2$. Therefore we can compute the mutual information between $C$ and $D$ :
\begin{IEEEeqnarray*}{rCl}
I(C\lon D) &=& \h(C) - \h(C|D) \\
   &=& \log{q^2} -  \frac{q(q-1)}{2}\frac2{q^2}\log\frac{q^2}{2} - q\frac1{q^2}\log{q^2}\\  
   &=& \log{q^2} -  \frac{q-1}{q}\log\frac{q^2}{2} - \frac1{q}\log{q^2}\\
   &=& \frac{q-1}{q}.
\end{IEEEeqnarray*}
When $A$ and $B$ are known and $A\not=B$, then $C$ is uniquely defined
(the only line incident with both points). If $A=B$ (which happens with probability
$1/q$) we need $\log q$ bits to specify $C$. Hence,
$$
\h(C|A,B) = \frac{\log q}{q}.
$$

Given $B$, the point  $A$ is either equal to $B$ or $A$ belongs to the set of $q(q-1)$ points that do not share the same abscissa as $B$.
$A = B$ happens for $q(q-1)$ elementary events: each of the $q$ lines going through $B$ are the tangents of $q-1$ parabolas.
$A\neq B$ happens for $q-1$ elementary events: there is one line through $A$ and $B$ and $q-1$ parabolas, and since there are $q(q-1)$ such points, we have
\begin{IEEEeqnarray*}{rCl}
I(A\lon B) &=& \h(A) - \h(A|B) \\
   &=& \log{q^2} -  q(q-1)\frac1{q^2}\log{q^2}  - \frac1{q}\log{q}\\  
   &=& 2\log{q} -  2\frac{q-1}{q}\log{q} - \frac1{q}\log{q} \\  
   &=&  \frac{\log q}{q}.\\
\end{IEEEeqnarray*}
Now we compute both sides of the inequality
$$
I(C\lon D) \le I(C\lon D|A) + I(C\lon D|B) + 
  \lambda_1 I(A\lon B) + \lambda_2 I(A\lon B|C) + \lambda_3 \h(C|AB)
$$
and get
$$
1 - \frac1{q} \le  \lambda_1\frac{\log q}{q}+\lambda_3\frac{\log q}{q}.
$$
This leads to a contradiction for large $q$. It follows that ($\I$) and ($\III$)
are essentially conditional.

\subsection{Stronger form of Theorem~\ref{th-not-unconditional} for inequalities ($\I$) and ($\III$)}

The construction in Section~\ref{geometric-counterexample} implies something
stronger than Theorem~\ref{th-not-unconditional} for inequalities ($\I$) and ($\III$).
Indeed, next proposition  follows from ($\I$) and from ($\III$):
\begin{eqnarray}
\label{ineq:weakform}
I(A\lon B|C) = I(A\lon B|D) = \h(C|A,B) = I(C\lon D|A)= 
I(C\lon D|B) =  I(A\lon B)=0 \Ra I(C\lon D) \le 0.
\end{eqnarray}
We claim that this conditional (in)equality is  also \emph{essentially conditional}:
\begin{theorem}\label{cond-eq}
Inequality~(\ref{ineq:weakform}) is essentially conditional, \ie for any values $(\lambda_i)_{1\le i\le 6}$, the inequality
$$
I(C\lon D) \le \lambda_1 I(C\lon D|A) + \lambda_2I(C\lon D|B) + 
\lambda_3I(A\lon B) + \lambda_4 I(A\lon B|C) + \lambda_5I(A\lon B|D) + \lambda_6\h(C|A,B)
$$

is not valid.
\end{theorem}
\begin{IEEEproof}
For  the quadruple $(A,B,C,D)_q$ from the geometric example defined 
in Section~\ref{geometric-counterexample},
each term in the right-hand side of the inequality vanishes as $q$ tends to
infinity, but the left-hand side does not.
\end{IEEEproof}
\begin{remark}
Conditional inequality (\ref{ineq:weakform}) is implied by either ($\I$) or ($\III$). Hence, 
Theorem~\ref{th-not-unconditional} is a corollary of Theorem~\ref{cond-eq}, \ie
 Theorem~\ref{cond-eq} is somewhat stronger than  Theorem~\ref{th-not-unconditional}.
\end{remark}

\section{Inequalities for Almost Entropic Points}\label{section-aep}

Let us recall that a point in $\mathbb{R}^{2^n-1}$ is called \emph{entropic} if
it represents the entropy profile of some distribution; a point
is called \emph{almost entropic} if it belongs to the closure of the set
of entropic points.  It is known that for every $n$ the set of almost
entropic points is a convex cone in $\mathbb{R}^{2^n-1}$, see \cite{yeung-book}. The
set of all almost entropic points is exactly the set of points that satisfy all unconditional
linear information inequalities.  Though some almost entropic points do
not  correspond to any distribution, we will abuse notation and refer to
coordinates of an almost entropic point as $\h(X_i)$, $\h(X_i,X_j)$, etc.
Moreover, we  keep the notation $I(X_i\lon X_j)$, $I(X_i\lon X_j|X_k)$, etc.
for the corresponding linear combinations of coordinates of these points. So we can naturally apply each linear information inequalities to the almost entropic points. The question is which conditional inequality remain valid for all almost entropic points and which do not.

\subsection{Some conditional inequalities fail for almost entropic points}\label{sub:section-aepoint}

We have seen that ($\IV$--$\VI$) hold for almost entropic points.
This is not the case for two other essentially conditional inequalities.
\begin{theorem}\label{th-counterexample-aep}
Inequalities  ($\I$) and ($\III$) do not hold for almost entropic points.
\end{theorem}
\begin{IEEEproof}
The main technical tool used in our proof is Slepian\textendash{}Wolf coding. We do not need the general
version of the classic Slepian\textendash{}Wolf theorem, we use only its special case  (actually this
special  case makes the most important part of the general proof of the standard Slepian\textendash{}Wolf
theorem, see Theorem~2 in the original paper  \cite{slepian-wolf} and a detailed 
discussion in section~5.4.4 of  \cite{cover-thomas}).
\begin{lemma}[Slepian\textendash{}Wolf coding]\label{lemma-sw}
Let $(X^m,Y^{m})$ be $m$ independent copies of jointly distributed random variables $(X,Y)$, \ie $X^m=(X_1,\ldots,X_m)$,
$Y^m=(Y_1,\ldots,Y_m)$, where pairs $(X_i,Y_i)$ are i.i.d. 
Then there exists $X'$ such that
\begin{itemize}
\item $\h(X'|X^m) = 0$,
\item $\h(X') = \h(X^m|Y^m) + o(m)$,
\item $\h(X^m|X',Y^m) = o(m)$.
\end{itemize}
\end{lemma}
(This  Lemma is also a special case of Theorem~3 in \cite{matus-two-constructions}.)
Lemma~\ref{lemma-sw} claims that we can construct a hash of a random variable $X$ which is almost
independent of $Y$ and has approximately the entropy of $X$ given $Y$. We
will say that $X'$ is the Slepian\textendash{}Wolf hash of $X^m$ given $Y^m$ and write $X' =
SW(X^m|Y^m)$.

\emph{Construction of an almost entropic counterexample for~($\I$)}:

\begin{enumerate}

\item
Start with distribution $(A,B,C,D)_q$ defined in Section~\ref{geometric-counterexample}. 
The value of $q$ is specified in what follows. For this distribution $I(A\lon B|C)=0$  but
$I(A\lon B)>0$. So far, distribution  $(A,B,C,D)_q$ does not satisfy the 
conditions of~($\I$). 

\item Serialize it: we define a new quadruple $(A^m,B^m,C^m,D^m)$ such that each entropy is
$m$ times greater. More precisely, distribution $(A^m,B^m,C^m,D^m)$ is obtained by sampling $m$ times independently
$(A_i,B_i,C_i,D_i)$ according to the distribution $(A,B,C,D)$ and letting
$A^m=(A_1,\ldots,A_m)$, $B^m=(B_1,\ldots,B_m)$, $C^m=(C_1,\ldots,C_m)$,
and $D=(D_1,\ldots,D_m)$.

\item Apply Slepian\textendash{}Wolf coding Lemma and define $A'=SW(A^m|B^m)$,
then replace  in the quadruple $A^m$ by $A'$.

The entropy profile for $A',B^m,C^m,D^m$ cannot be far different from
the entropy profile for $A^m,B^m,C^m,D^m$.  Indeed, by construction, 
$\h(A'|A^m)=0$ and 
\begin{align*} 
\h(A^m|A') 
&= \h(A^m)+\h(A'|A^m)- \h(A')\\
&= \h(A^m) - \h(A^m|B^m) + o(m)\\
&= I(A^m\lon B^m)+o(m).
\end{align*}
Hence, the difference
between entropies involving $(A',B^m,C^m,D^m)$ and $(A^m,B^m,C^m,D^m)$
is at most 
$I(A^m\lon B^m) + o(m) = O\left(\frac{\log q}{q}m\right)$. 

Notice that $I(A'\lon B^m|C^m)=0$ since $A'$ functionally depends on $A^m$ and 
in the initial distribution $I(A^m\lon B^m|C^m) = 0$.

\item Scale down the entropy profile of $(A',B^m,C^m,D^m)$ by a factor of $1/m$. 
More precisely, we use the following fact. If the entropy profile of $(A', B^m, C^m, D^m)$ is
some point $\vec h\in\mathbb{R}^{15}$, then for every $\varepsilon>0$
there exists another 
distribution  $(A'',B'',C'', D'')$ with an entropy profile $\vec h'$ such that 
$
\|\vec h' - \frac{1}{m} \vec h\| < \varepsilon. 
$
This follows from  convexity of the set 
of almost entropic points. (The new distribution can be constructed
explicitly,  see Theorem~14.5 in \cite{yeung-book}). 
We may assume that  $\varepsilon=1/m$.

\item Tend $m$ to infinity. 
The resulting entropy profiles tend to some limit,
which is an almost entropic point. 
This point does not satisfy
($\I$)  (for $q$ large enough). Indeed, on one hand, 
the values of $I(A''\lon B'')$ and $I(A''\lon B''|C'')$ 
converge to zero. On the other hand,  
inequality $I(C''\lon D'') \le I(C''\lon D''|A'') + I(C''\lon D''|B'')$ results  in
$$
1-O\left(\frac{\log_2 q}{q}\right) \le 
O\left(\frac{\log_2 q}{q}\right),
$$
which can not hold for large enough  $q$.
\end{enumerate}
\medskip

\emph{Construction of an almost entropic counterexample for Inequality ($\III$):} 
In this construction we need another lemma based on Slepian\textendash{}Wolf coding.
\begin{lemma}
\label{rel-lemma}
For every distribution $(A,B,C,D)$ and every integer $m$ there exists a
distribution $(A',B',C',D')$ such that the following three conditions hold.
\begin{itemize}
\item $\h(C'|A', B')=o(m)$.
\item Denote $\vec h$ the entropy profile  of $(A,B,C,D)$ and $\vec h'$ the
entropy profile of $(A',B',C',D')$; then the components of $\vec h'$ differ from
the corresponding components of $m\cdot \vec h$ by at most $m\cdot \h(C|A,B) +
o(m)$.
\item Moreover, if  in the original distribution $I(A\lon B|C) = 0$, then 
$I(A'\lon B'|C') = o(m)$.
\end{itemize}
\end{lemma}
(This lemma is proven in Appendix.) Now we construct an almost entropic counterexample to ($\III$).

\begin{enumerate}

\item Start with the distribution $(A,B,C,D)_q$ 
from Section~\ref{geometric-counterexample} (the value of $q$ is chosen later).


\item Apply Lemma~\ref{rel-lemma} and get
$(A',B',C',D')$ such that $\h(C'|A',B')=o(m)$. Lemma~\ref{rel-lemma} guarantees that other
entropies of $(A',B',C',D')$  are about $m$ times larger then the corresponding entropies for
$(A,B,C,D)$, possibly with an overhead of size $$O(m\cdot \h(C|A,B)) =
O\left(\frac{\log_2 q}{q}m\right).$$ 
From the last bullet of Lemma~\ref{rel-lemma} we also get 
$I(A'\lon B'|C') = o(m)$. 
 
\item Scale down the entropy point of $(A',B',C',D')$ by the factor of $1/m$
within precision of $1/m$, similarly to step (4) in the previous
construction.

\item Tend $m$ to infinity to get an almost entropic point. Conditions of
($\III$) are satisfied for $I(A'\lon B'|C')$ and $ \h(C'|A', B')$ both
vanish in the limit. Inequality ($\III$)  reduces to
$$
1-O\left(\frac{\log_2 q}{q}\right) \le 
O\left(\frac{\log_2 q}{q}\right),
$$
which can not hold if $q$ is large enough.
\end{enumerate}
\end{IEEEproof}

The proven results can be rephrased as follows: There exist almost entropic
points that satisfy all unconditional linear inequalities for entropies
but do not satisfy conditional inequalities ($\I$) and $(\III)$ respectively.

Note that one single (large enough) value of $q$ suffices to construct
almost entropic counterexamples for ($\I$) and $(\III)$. However the
choice of $q$ in the construction of Theorem~\ref{th-counterexample-aep}
provides some freedom: we can control the gap between the left-hand side and
the right-hand side of the inequalities. By increasing $q$ we can make 
the ratio between the left-hand side and the right-hand side of Inequalities ($\I$) and $(\III)$ 
greater than any assigned number.

\section{The cone of almost entropic points is not polyhedral\label{sub:section-inf-inequalities}}

In section~\ref{subsection-geometric-meaning} we discussed the geometric meaning of essentially conditional inequalities and claimed that the existence of an essentially conditional inequality that is valid for almost entropic points implies that the cone of almost entropic points is not polyhedral. Now present a formal proof of this result. 

Our proof is based on the following classic lemma widely used in  linear and convex programming.
\begin{lemma}[Farkas' lemma] 
Let 
$
g(\mathbf{x}), h_1(\mathbf{x}), \ldots, h_s(\mathbf{x})
$
be  real-valued linear functionals on $\mathbb{R}^n$. Then, exactly one of the following two statements is true:
\begin{itemize}
\item[(a)] There exists a tuple of non-negative real numbers $(\kappa_1,\ldots,\kappa_s)$  
such that
$$
g(\mathbf{x}) =  \kappa_1 h_1(\mathbf{x})+ \ldots + \kappa_s h_s(\mathbf{x}),
$$
\ie  $g$ can be represented as a linear combination of  $h_j$ with non-negative coefficients.
\item[(b)] There exists an $\mathbf{x}\in\mathbb{R}^n$ such that 
$$
 h_1(\mathbf{x}) \ge 0, \ldots, h_s(\mathbf{x})\ge 0,
$$
and $g(\mathbf{x})<0$.
\end{itemize}
\end{lemma}
We will use the nontrivial direction of this alternative: if (b) is false than (a) is true.
Also we will need the following simple lemma from linear algebra
(see, e.g., \cite[Theorem~2 in Section~17]{halmos}
or \cite[Theorem~4.7]{rudin}):
\begin{lemma}\label{lemma-double-duality}
Let $f_i, \ i=1,\ldots m$ be a family of linear functionals on a finite-dimensional real linear space, and
$\cal L$ be the set (the linear subspace) of all vectors on which every $f_i$ vanishes, \ie
 $$
 \mathbf{x}\in {\cal L} \Leftrightarrow f_1(\mathbf{x})=\ldots=f_m(\mathbf{x})=0.
 $$
Assume that another linear functional $g(\mathbf{x})$ is identically equal to $0$ for all vectors in $\cal L$.
Then $g$ is a linear combination of functionals $f_i$.
\end{lemma}

\begin{theorem}\label{th-non-polyhedral-general}
Let $\cal C$ be a polyhedral closed convex cone in $\mathbb{R}^N$ and~$ f_1, \ldots, f_m$ and $g$
be linear functionals on  $\mathbb{R}^N$.
Assume that the conditional inequality
\begin{equation}
\label{general-conditional-ineq}
(f_i(\mathbf{x}) = 0 \mbox{ for all }i=1,\ldots,m) \Ra g(\mathbf{x}) \ge 0
\end{equation}
holds for all $\mathbf{x}\in {\cal C}$. Then there exist real numbers  $(\lambda_i)_{1\le i\le m}$
such that  the inequality
\begin{equation}\nonumber
g(\mathbf{x}) + 
\sum\limits_{i=1}^m{\lambda_i f_i(\mathbf{x})} \ge 0 
\end{equation}
holds for all $\mathbf{x}\in  {\cal C}$.  (In a word,
for a polyhedral closed convex cone every conditional linear inequality is \emph{not} essentially conditional;
\emph{cf}.~a $2$-di\-men\-si\-o\-nal example in Fig.~1.)
\end{theorem}
\begin{IEEEproof} 
Since $ {\cal C}$ is polyhedral,  there exists a finite set of linear functionals $h_j$, $j=1,\ldots,s$ such that 
$$
\mathbf{x}\in  {\cal C} \Leftrightarrow (h_j(\mathbf{x})\ge 0 \mbox{ for all } j=1,\ldots,s).
$$
Denote by ${\cal L}$ the linear subspace on which  all $f_i$ vanish, \ie 
$$
\mathbf{x}\in {\cal L}  \Leftrightarrow f_1(\mathbf{x})=\ldots=f_m(\mathbf{x})=0.
$$
We are given that for every  $\mathbf{y} \in {\cal L}$
such that $h_j(\mathbf{y})\ge 0$ for $j=1,\ldots,s$ (\ie for every $\mathbf{y} \in {\cal L}\cap  {\cal C}$) it holds  that $g(\mathbf{y}) \ge 0$. We apply Farkas'  lemma to the linear subspace $\cal L$  and conclude that for some  reals $\kappa_j\ge 0$
\begin{eqnarray}\label{alpha-prim}
  g(\mathbf{y}) =  \kappa_{1} h_1(\mathbf{y}) + \ldots +\kappa_s h_s(\mathbf{y})
\end{eqnarray}
for all $\mathbf{y} \in {\cal L}$. In other words, the functional
\begin{eqnarray}\nonumber
   \kappa_{1} h_1 (\mathbf{x}) + \ldots +\kappa_s h_s(\mathbf{x})  - g(\mathbf{x}) 
\end{eqnarray}
is identically equal to $0$ for all vectors  $\mathbf{x}\in{\cal L}$.
Now we extend this identity from $\cal L$ to the entire $\mathbb{R}^n$.
From  Lemma~\ref{lemma-double-duality} it follows that for some reals $\lambda_j$ the equality
\begin{eqnarray*}
  \kappa_{1} h_1 (\mathbf{x}) + \ldots +\kappa_s h_s(\mathbf{x})  - g(\mathbf{x})
 {=}  \lambda_1 f_1(\mathbf{x}) + \ldots+ \lambda_m f_m(\mathbf{x})
\end{eqnarray*}
holds for all $\mathbf{x}\in\mathbb{R}^n$.  

Each functional $h_j(\mathbf{x})$  is non-negative for all $\mathbf{x}\in  {\cal C}$. 
Hence,  for every $\mathbf{x}\in  {\cal C}$
$$
 g(\mathbf{x})  + \lambda_1 f_1(\mathbf{x}) + \ldots + \lambda_m f_m(\mathbf{x})\ge0,
$$
 and we are done.
\end{IEEEproof}

From   Theorem~\ref{th-not-unconditional} and Theorem~\ref{th-non-polyhedral-general} we  conclude immediately  the the cone of almost entropic points in not polyhedral:

\begin{theorem}[F.~Mat\'{u}\v{s}, \cite{Matus-inf}] \label{th-non-polyhedral}
For $n\ge 4$  the cone of almost entropic points for distributions of $n$-tuples
of random variables
is not polyhedral. Equivalently, the cone of (unconditional) linear inequalities for
the entropies of $4$-tuples of random variables is not  polyhedral.
\end{theorem}
\begin{IEEEproof}
On the one hand, for every $n$ the set of all almost entropic points for $n$-tuples of random variables
is a closed convex cone, see \cite{yeung-book}.
On the other hand, for all $n\ge 4$ for the cone of almost entropic points
there exist essentially conditional inequalities (with linear constraint that specify some ``facet'' of this cone). 
Indeed, we can employ  ($\IV'$) or ($\V'$); by Theorem~\ref{th-matus-07} these inequalities hold for all almost entropic points, and by Theorem~\ref{th-not-unconditional} they are essentially conditional.
Hence, from Theorem~\ref{th-non-polyhedral-general}  it follows that the cone of almost entropic points cannot be polyhedral.
\end{IEEEproof}
\begin{remark}
The original proof of this theorem in  \cite{Matus-inf}  is analogous to the argument above,  instantiated with Inequality~($\V'$).
\end{remark}

\section{Conditional inequalities for Kolmogorov complexity\label{section-kolm}}


The Kolmogorov complexity of a finite binary string $\word{x}$ is defined as the length of
a shortest program that prints $X$ on the empty input; 
similarly, the conditional Kolmogorov complexity of a
string $\word{x}$ given another string $\word{y}$ is defined as the length of a shortest program
that prints $\word{x}$ given $\word{y}$ as an input. More formally, for any programming
language\footnote{Technically a \emph{programming language} is  a partial computable function of two arguments,
${\cal L}\lon \{0,1\}^*\times \{0,1\}^*\to\{0,1\}^*$, which maps a pair consisting of program $p$ and its input $y$ to the output $x$ (the result printed by program $p$ given input $x$). If the value ${\cal L}(p,y)$ is undefined we say that  program $p$ does not halt when given input $y$.} $\cal L$, the Kolmogorov complexity $\C_{\cal L}(\word{x}|\word{y})$ is defined as
 $$
\C_{\cal L}(\word{x}|\word{y}) = \min\{|p| : \mbox{ program $p$ prints $\word{x}$ on input $\word{y}$}\},
 $$
and unconditional complexity $\C_{\cal L}(\word{x})$ is defined as complexity of $\word{y}$ 
given the empty $\word{y}$. 
The basic fact of Kolmogorov complexity theory is the invariance
theorem: there exists a universal programming language $\cal U$ such that for any
other language $\cal L$ we have $\C_{\cal U}(\word{x}|\word{y})  \le \C_{\cal L}(\word{x}|\word{y})+O(1) $ 
(the $O(1)$ term here depends on $\cal L$ but not on $\word{x}$ and $\word{y}$). 
We fix such a universal language $\cal U$; in what
follows we omit the subscript $\cal U$ and denote the Kolmogorov complexity by $\C(\word{x})$,
$\C(\word{x}|\word{y} )$. We refer the reader to \cite{lv} for an exhaustive survey on
Kolmogorov complexity.

We introduce notation for  the complexity profile of a tuple of strings
similar to the notation for entropy profile  (defined in section~\ref{sub:section-notation}). 
We fix some computable encoding of finite tuples of binary strings (a computable bijection between
the set of all finite tuples of binary strings and the set of  individual binary strings). Now we can talk
about Kolmogorov complexity of pairs of strings, triples of strings, etc.: Kolmogorov complexity of 
a tuple is defined as Kolmogorov complexity of its code.
Let $\mathcal{X} =( \word{x}_1,\ldots,\word{x}_n)$ be $n$ binary strings;
then Kolmogorov complexity is associated to each  subtuple $( \word{x}_{i_1},\ldots,\word{x}_{i_k})$, and
we define   \emph{complexity profile} of $\mathcal{X}$  as
$$\vec{\C}(\mathcal{X}) = (\C(\word{x}_{i_1},\ldots,\word{x}_{i_k}))_{1\le i_1<\cdots<i_k\le n},$$
\ie a vector of  complexities of $2^n-1$ non-empty tuples in the lexicographic order. 
We also need a similar notation for the vector of conditional complexities: 
if $\mathcal{X} =( \word{x}_1,\ldots,\word{x}_n)$ is an $n$-tuple of  binary strings and $y$ is another binary string,
then we denote
$$\vec{\C}(\mathcal{X}|\word{y}) = (\C(\word{x}_{i_1},\ldots,\word{x}_{i_k}|\word{y}))_{1\le i_1<\cdots<i_k\le n}.$$

Kolmogorov complexity was introduced in \cite{kol65} as an algorithmic version of
Shannon's measure of information. It is natural to expect that the basic
properties of
Kolmogorov complexity are  similar in some sense to the properties of Shannon entropy. 
We have indeed many examples of parallelism between Shannon entropy and
Kolmogorov complexity. For example, for the property of Shannon entropy 
$\h(X,Y) = \h(X) + \h(X|Y)$ (where $X$ and $Y$ are random variables) 
there is a  counterpart in the Kolmogorov setting
$\C(\word{x}, \word{y}) = \C(\word{x}) + \C(\word{x}|\word{y}) + O(\log (|\word{x}|+|\word{y}|))$ (where $\word{x}$ and $\word{y}$ are binary strings).
This statement for Kolmogorov complexity is called the Kolmogorov--Levin theorem,
see \cite{zl}. This result justifies the definition of the
mutual information, which is an algorithmic version of  Shannon's standard
definition: the mutual information between binary strings can be defined as 
 $$
 \info(\word{x} \lon \word{y}) := \C(\word{x}) + \C(\word{y}) - \C(\word{x}, \word{y}), 
 $$
and the conditional mutual information can be defined as
 $$
 \info(\word{x} \lon \word{y}|\word{z}) := \C(\word{x}, \word{z}) + \C(\word{y}, \word{z}) - \C(\word{x}, \word{y}, \word{z}) - \C(\word{z}).
 $$
From the Kolmogorov\textendash{}Levin theorem it follows that $\info(\word{x} \lon \word{y})$ is equal to
$\C(\word{x})-\C(\word{x}|\word{y})$, and the conditional mutual information $\info(x \lon y|z)$ is equal to
$\C(\word{x}|\word{z}) - \C(\word{x}|\word{y}, \word{z})$ (all these equalities hold only up to additive logarithmic terms).

Actually a much more  general similarity between Shannon's
and Kolmogorov's information theories is known. For every linear inequality for
Shannon entropy there exists a counterpart for Kolmogorov complexity.
\begin{theorem}[Inequalities are the same, \cite{hrsv}]\label{th-hrsv}
 For each family of real coefficients $\{\kappa_W\}$ the inequality
  $$
  \sum\limits_{i}\kappa_i \h(X_i) +\sum\limits_{i<j}\kappa_{i,j} \h(X_i,X_j)+\ldots \ge 0
  $$
is true for every distribution $(X_i)$  if and only if for some $\const>0$
the inequality
  $$
  \sum\limits_{i}\kappa_i \C(\word{x}_i) +\sum\limits_{i<j}\kappa_{i,j} \C(\word{x}_i,\word{x}_j)+\ldots +\const\cdot  \log n\ge 0
  $$
is true for all tuples of strings $(\word{x}_i)$, where $n$ denotes the sum of the lengths of all strings $x_i$ and
 $\const$ does not depend on strings $\word{x}_i$.
 \end{theorem}
Thus, the class of unconditional inequalities valid for Shannon entropy
coincides with the class of (unconditional) inequalities valid for Kolmogorov
complexity. 
 \begin{remark}\label{hrsv-constant}
An analysis of the proof in \cite{hrsv} shows that for a fixed size of the tuple (a fixed number of random variables/binary strings involved in inequalities) the value of $\const$ in Theorem~\ref{th-hrsv} can be bounded by $O\left(\sum \kappa_W\right)$.
 \end{remark}

Can we extend this parallelism to the conditional information inequalities?  It
is not obvious how to even formulate conditional inequalities for Kolmogorov
complexity.  A subtle point is that in the framework of Kolmogorov complexity we
cannot say that some information quantity exactly equals zero. Indeed, even the
definition of Kolmogorov complexity makes sense only up to an additive term
that depends on the choice of the universal programming language.  Moreover,
such a natural basic statement as the Kolmogorov\textendash{}Levin theorem holds only up
to a logarithmic term. It makes no sense to say that $\info(\word{a}\lon \word{b})$ or $\info(\word{a}\lon
\word{b}|\word{c})$  \emph{exactly} equals zero, we can only  require that these information
quantities are negligible comparative to the lengths of strings $\word{a},\word{b},\word{c}$.  So,
if we want to prove a meaningful conditional inequality for Kolmogorov
complexity, the linear constraints for information quantities must be
formulated with some reasonable  precision.

In this section we prove two different results. One of these results (Theorem~\ref{th-matus-07-kolm}) is positive. It shows that some  counterparts of inequalities ($\IV$--$\VI$) hold for Kolmogorov complexity. The other result of this section (Theorem~\ref{th-counterexample-kolm}) is negative. It claims that natural counterparts of inequalities ($\IV$) and ($\VI$) do not hold for Kolmogorov complexity.

We also want to mention a peculiar property of the conditional inequalities that we get in Theorem~\ref{th-matus-07-kolm}. These two inequalities are not valid with the logarithmic precision (sharply contrast to all  unconditional linear inequalities for Kolmogorov complexity).  In fact, if we take the linear constraints with a precision of $O(\log n)$, then the resulting conditional inequalities are valid only up to $O(\sqrt{n\log n})$.

\subsection{Positive result: three conditional inequalities make sense for Kolmogorov complexity}

Now we show that inequalities from Theorem~\ref{th-matus-07} can be translated in some sense
in the language of Kolmogorov complexity.  In this section we use for quadruples of strings $\word{a},\word{b},\word{c},\word{d}$ the notation
$\square_{\word{a}\word{b},\word{c}\word{d}} := \info(\word{c}\lon \word{d}|\word{a}) + \info(\word{c}\lon \word{d}|\word{b}) +  \info(\word{a}\lon \word{b}) -  \info(\word{c}\lon \word{d})$,
similar to the box notation used for Shannon entropy (introduced in Section~\ref{sub:section-notation}).
\begin{theorem}\label{th-matus-07-kolm} Let $f(n)$ be a function of an integer argument such that $f(n)\ge \log n$ for all $n$. Then there exists a number $\const>0$ such that
for every tuple of binary strings $(\word{a},\word{b},\word{c},\word{d},\word{e})$ 

\smallskip

\noindent
$(\IV\mbox{-Kolm})$\ if $\info(\word{a}\lon \word{d}|\word{c})\le f(n)$ and $\info(\word{a}\lon \word{c}|\word{d})\le f(n)$, then
 $\square_{\word{a}\word{b},\word{c}\word{d}} + \info(\word{a}\lon \word{c}|\word{e})+\info(\word{a}\lon \word{e}|\word{c})+\const\cdot  \sqrt{n\cdot f(n)}\ge0$, 

\medskip

\noindent
$(\V\mbox{-Kolm})$\  if $\info(\word{b}\lon \word{c}|\word{d})\le f(n)$ and  $\info(\word{c}\lon \word{d}|\word{b})\le f(n)$, then
$\square_{\word{a}\word{b},\word{c}\word{d}} +  \info(\word{b}\lon \word{c}|\word{e})+\info(\word{c}\lon \word{e} |\word{b})+\const\cdot \sqrt{n\cdot f(n)}\ge0$,

\medskip

\noindent
$(\VI\mbox{-Kolm})$\ if $\info(\word{b}\lon \word{c}|\word{d})\le f(n)$  and $\info(\word{c}\lon \word{d}|\word{b})\le f(n)$, then
$\square_{\word{a}\word{b},\word{c}\word{d}} + \info(\word{c}\lon \word{d}|\word{e})+\info(\word{c}\lon \word{e}|\word{d})+\const\cdot \sqrt{n\cdot f(n)}\ge0$,

\smallskip

\noindent
where $n$ is the sum of lengths of strings $\word{a},\word{b},\word{c},\word{d},\word{e}$.
\end{theorem}
In this theorem $f(n)$ plays the role of the measure of ``precision'' of the constraints.
For example, assuming $\info(\word{a}\lon \word{d}|\word{c}) = O\left(\sqrt{n}\right)$ and $\info(\word{a}\lon \word{c}|\word{d})=O\left(\sqrt{n}\right)$ we get from the theorem that there exists a $\theta>0$ such that for all 
$\word{a},\word{b},\word{c},\word{d},\word{e}$
$$\square_{\word{a}\word{b},\word{c}\word{d}} + \info(\word{a}\lon \word{c}|\word{e})+\info(\word{a}\lon \word{e}|\word{c}) +\theta\cdot n^{3/4} \ge 0.$$
\begin{IEEEproof}
By Theorem~\ref{th-hrsv},
for every linear inequality for Shannon entropy there exists a counterpart for Kolmogorov
complexity that is true for all binary strings up to an additive $O(\log n)$-term. Thus, 
from Inequality ($i$) on page~\pageref{uncond-ineq} (which holds for the Shannon entropies of any distribution) 
it follows that a similar inequality holds for Kolmogorov complexity. More precisely, for each integer $k>0$ there exists a real $\const'=\const'(k)$ such that for all strings $\word{a},\word{b},\word{c},\word{d},\word{e}$
\begin{eqnarray}
\begin{aligned}
\label{eq:kolm1}
\square_{\word{a}\word{b},\word{c}\word{d}}+ \info(\word{a}\lon \word{c}|\word{e})+\info(\word{a}\lon \word{e}|\word{c}) +\frac{1}{k}\info(\word{c}\lon \word{e}|\word{a}) + 
\frac{k-1}{2}(\info(\word{a}\lon \word{d}|\word{c})+\info(\word{a}\lon \word{c}|\word{d})) + \const'\cdot \log n\ge 0.
\end{aligned}
\end{eqnarray}
By Remark~\ref{hrsv-constant}  we may assume that $\const'(k)=O(k)$.

It remains to choose $k$ that makes the left-hand side of~(\ref{eq:kolm1}) as small as possible
(making this inequality as strong as possible). The value of $k$ can depend on strings $\word{a},\word{b},\word{c},\word{d}$. More technically,  for every $n$ (the sum of lengths of the strings involved in the inequality) we choose a suitable $k=k(n)$.

The value $\frac{1}{k}\info(\word{c}\lon \word{e}|\word{a})$ is bounded by $O(n/k)$ since all strings are of length
at most $n$;  the values $\info(\word{a}\lon \word{d}|\word{c})$ and $\info(\word{a}\lon \word{c}|\word{d})$ are less than $f(n)$;  the term $(\const'\cdot \log n)$ is dominated  by $O(k\cdot f(n))$.
Hence, Inequality~(\ref{eq:kolm1}) rewrites to
$$
-\square_{\word{a}\word{b},\word{c}\word{d}}- \info(\word{a}\lon \word{c}|\word{e})-\info(\word{a}\lon \word{e}|\word{c})  \le O(n/k)+O(k \cdot f(n)).
$$
Thus, we should choose  $k$  that  minimizes the sum of two term $O(n/k)+O(k\cdot f(n))$. By letting $k=\sqrt{n/f(n)}$, we make these two terms equal to each other up to a constant factor; so their sum becomes minimal (also up to a constant factor).
For the chosen $k$ Inequality~(\ref{eq:kolm1})  results in
 $$-\square_{\word{a}\word{b},\word{c}\word{d}} - \info(\word{a}\lon \word{c}|\word{e})-\info(\word{a}\lon \word{e}|\word{c})\le O(\sqrt{n\cdot f(n)}),$$ 
 and ($ \IV\mbox{-Kolm}$) is proven.  Conditional inequalities ($\V\mbox{-Kolm}$) and ($\VI\mbox{-Kolm}$) can be proven by a similar argument. 
\end{IEEEproof}

Theorem~\ref{th-matus-07-kolm} involves $5$-tuples of strings $(\word{a},\word{b},\word{c},\word{d},\word{e})$ but
it implies a nontrivial result for quadruples of strings. 
By assuming $\word{e}=\word{d}$ we get from Theorem~\ref{th-matus-07-kolm} the following corollary.
\begin{corollary}\label{cor-matus-07-kolm}  Let $f(n)$ be a function of an integer argument such that $f(n)\ge \log n$ for all $n$.
Then there exists a $\theta>0$ such that for every tuple of binary strings $(\word{a},\word{b},\word{c},\word{d})$
$$
\begin{array}{rll}
(\IV'\mbox{-Kolm}) & \mbox{if } \info(\word{a}\lon \word{d}|\word{c})\le f(n) \mbox{ and } \info(\word{a}\lon \word{c}|\word{d})\le f(n), 
 \mbox{ then } \square_{\word{a}\word{b},\word{c}\word{d}} +\theta\cdot\sqrt{n\cdot f(n)},\\
 \\
(\V'\mbox{-Kolm}) & \mbox{if }  \info(\word{b}\lon \word{c}|\word{d})\le f(n) \mbox{ and } \info(\word{c}\lon \word{d}|\word{b})\le f(n),
\mbox{ then } \square_{\word{a}\word{b},\word{c}\word{d}} +\theta\cdot\sqrt{n\cdot f(n)}),
\end{array}
$$
\noindent
where $n$ is the sum of the lengths of all strings involved.
\end{corollary}

In Theorem~\ref{th-matus-07-kolm} and Corollary~\ref{cor-matus-07-kolm} we deal with two different 
measures of precision: $f(n)$ in the conditions and $O(\sqrt{n \cdot f(n)})$ 
in the conclusions. These two measures of precision can be  dramatically different. 
Assume, for example, that  $\info(\word{b}\lon \word{c}|\word{d})$  and $\info(\word{c}\lon \word{d}|\word{b})$ are bounded by $O(\log n)$,
which is the most natural conventional assumption of ``independence'' in algorithmic information theory.
Then  from Corollary~\ref{th-matus-07-kolm} it follows only that $ \square_{\word{a}\word{b},\word{c}\word{d}} +\theta\cdot \sqrt{n\log n}\ge0$.
Can we prove the same inequality with a precision better than $O(\sqrt{n\log n})$? The answer is negative: the next proposition shows that this bound is tight.

\begin{proposition}\label{prop-lower-bound-k-complexity}
For some $\const>0$, 
for infinitely many integers $n$ there exists a tuple of strings $(\word{a},\word{b},\word{c},\word{d})$ such that  $\C(\word{a},\word{b},\word{c},\word{d})=n+O(1)$,
 $\info(\word{b}\lon \word{c}|\word{d})=O(\log n)$  and $\info(\word{c}\lon \word{d}|\word{b})=O(\log n)$, and 
$$
\square_{\word{a}\word{b},\word{c}\word{d}} \le - \const\cdot\sqrt{n\log n}.
$$
\end{proposition}
\begin{IEEEproof}
Let us take the distribution   from Claim~5, p.~\pageref{claim5} for some
parameter $\varepsilon$, and denote it $(A,B,C,D)_\varepsilon$.
Further, we use the following  lemma from \cite{romash}.
\begin{lemma}[\cite{romash}]\label{lemma-fromh-to-c}
Let  $(A,B,C,D)$ be a distribution on some finite set $\mathcal{M}^4$, and $n$ be an integer. Then there exists a tuple of strings $(\word{a},\word{b},\word{c},\word{d})$ 
of length $n$ over alphabet $\mathcal{M}$
such that 
$
\vec \C(\word{a},\word{b},\word{c},\word{d}) = n\cdot \vec \h(A,B,C,D) + O(|\mathcal{M}|\log n).
$
\end{lemma}
\begin{remark}
The construction in the proof Lemma~\ref{lemma-fromh-to-c} is quite simple. For this particular distribution 
$(A,B,C,D)_\varepsilon$ we should take a quadruple of strings $(\word{a},\word{b},\word{c},\word{d})$
of length $n$ with frequencies of bits corresponding to the probabilities in the given distribution:
$$
\begin{array}{rcl}
 \#\{i\ :\ \word{a}_i=0, \word{b}_i=0, \word{c}_i=0,  \word{d}_i=0 \}
   &=& (\frac12-\varepsilon)n+O(1), \\
   \#\{i \ :\ \word{a}_i=0, \word{b}_i=1, \word{c}_i=0, \word{d}_i=1 \}
   &=& (\frac12-\varepsilon)n+O(1), \\
  \#\{i\ :\ \word{a}_i=1, \word{b}_i=0, \word{c}_i=1,  \word{d}_i=0 \}
   &=&  \varepsilon n+O(1), \\   
   \#\{i\ :\ \word{a}_i=1, \word{b}_i=1, \word{c}_i=0, \word{d}_i=0 \}
   &=&  \varepsilon n+O(1)
\end{array}
$$
(here $\word{a}_i$, $\word{b}_i$, $\word{c}_i$, and $\word{d}_i$ denote the $i$-th bit of the words $\word{a}$, $\word{b}$, $\word{c}$, $\word{d}$ respectively; $i=1,\ldots,n$).
More precisely, we should take a quadruple of binary strings with frequencies of bits as above with \emph{maximal possible} complexity $\C(\word{a},\word{b},\word{c},\word{d})$. It is proven in \cite{romash}
that for such a tuple of strings all Kolmogorov complexities involving $(\word{a},\word{b},\word{c},\word{d})$ are proportional to the corresponding Shannon entropies of $(A,B,C,D)_\varepsilon$
(up to an additive logarithmic term), \ie
 $$
\vec \C(\word{a},\word{b},\word{c},\word{d}) = n\cdot \vec \h(A,B,C,D) + O(\log n).
$$

\end{remark}
Thus, applying Lemma~\ref{lemma-fromh-to-c} to the distribution from Claim~5, p.~\pageref{claim5}
we get a tuple of binary strings $(\word{a},\word{b},\word{c},\word{d})$ such that
the quantities $\info(\word{c}\lon \word{d}|\word{a})$, $\info(\word{c} \lon \word{d}|\word{b})$, $\info(\word{a}\lon \word{b})$
are bounded by $O(\log n)$, while
$
\info(\word{c}\lon \word{d})= \Theta(\varepsilon n)
$ 
and 
$ 
\info(\word{b}\lon \word{c}|\word{d})=O(\varepsilon^2 n).
$

It remains to choose appropriate $\varepsilon$ and $n$. Let  $n$ be large enough and $\varepsilon = \sqrt{\frac{\log n}{n}}$.
Then $\info(\word{b}\lon \word{c}|\word{d}) = O(\varepsilon^2 n) = O(\log n)$ and 
$
\info(\word{c}\lon \word{d})\ge \const'\cdot \sqrt{n\log n}
$
(for some $\const'>0$).
Hence, we get
$$
\square_{\word{a}\word{b},\word{c}\word{d}} \le -\const'\cdot \sqrt{n\log n}+O(\log n),
$$
(the  multiplicative constant in the term $O(\log n)$ does not dependent on $\varepsilon$). So, for every $\const<\const'$ and for all large enough $n$
$$
\square_{\word{a}\word{b},\word{c}\word{d}} \le -\const\cdot \sqrt{n\log n},
$$
and we are done.
\end{IEEEproof}
\begin{remark}
Keeping in mind  details of the construction  from Claim~5, we can let  in the argument above $\const'=1$ 
and choose $\const$ arbitrarily close to $1$
(though the precise value of  constants does not much matter for 
Proposition~\ref{prop-lower-bound-k-complexity}).
\end{remark}

\subsection{Negative result: two conditional inequalities are not valid Kolmogorov complexity}

The following theorem shows that  counterparts of  ($\I$) and ($\III$) do not hold for Kolmogorov complexity.
\begin{theorem}\label{th-counterexample-kolm}
(a) There exists an infinite sequence of tuples of strings $(\word{a},\word{b},\word{c},\word{d})_n$ such that
the lengths of all strings $\word{a},\word{b},\word{c},\word{d}$ are $\Theta(n)$,
$\info(\word{a}\lon \word{b}) = O(\log n)$, $\info(\word{a}\lon \word{b} | \word{c}) = O(\log n)$, and 
$$
\info(\word{c}\lon \word{d}) - \info(\word{c}\lon \word{d}|\word{a}) - \info(\word{c}\lon \word{d}|\word{b}) = \Omega(n)
$$

(b) There exists an infinite sequence of tuples of strings $(\word{a},\word{b},\word{c},\word{d})_n$ such that
the lengths of all strings $\word{a},\word{b},\word{c},\word{d}$ are $\Theta(n)$,
$C(\word{c}|\word{a},\word{b}) = O(\log n)$, $\info(\word{a}\lon \word{b} | \word{c}) = O(\log n)$, and 
$$
\info(\word{c}\lon \word{d}) - \info(\word{c}\lon \word{d}|\word{a}) - \info(\word{c}\lon \word{d}|\word{b}) - \info(\word{a}\lon \word{b})= \Omega(n).
$$
\end{theorem}
\begin{remark}
Our proof of Theorem~\ref{th-counterexample-kolm}  is a transposition of the argument used in Theorem~\ref{th-counterexample-aep} from Shannon's theory setting to the setting of algorithmic information theory. 
The most important technical tool in the proof of Theorem~\ref{th-counterexample-aep} is
the Slepian\textendash{}Wolf theorem. We cannot apply it directly in the setting of Kolmogorov complexity.
Instead of it we  employ Muchnik's theorem on conditional descriptions (which is an algorithmic counterpart of the Slepian\textendash{}Wolf theorem):
\end{remark}
\begin{theorem}[Muchnik, \cite{muchnik}]\label{thm-muchnik}
For all strings  $\word{x},\word{y}$, there exists a string $\word{z}$ such that
\begin{itemize}
\item $|\word{z}| = \C(\word{x}|\word{y})$,
\item $\C(\word{z}|\word{x}) = O(\log n)$,
\item $\C(\word{x}|\word{y},\word{z}) = O(\log n)$,
\end{itemize}
where $n=\C(\word{x},\word{y})$.
We denote this string $\word{z}$ by $ Much(\word{x}|\word{y})$.
\end{theorem}

\begin{IEEEproof}[Proof of Theorem~\ref{th-counterexample-kolm}(a)]
We start with the distribution defined in Section~\ref{geometric-counterexample},
the value of $q$ is specified in what follows. Let us denote this distribution $(A,B,C,D)_q$.
We apply Lemma~\ref{lemma-fromh-to-c} and construct  a quadruple of strings $(\word{\hat a},\word{b},\word{c},\word{d})$ such that
$$
\vec \C(\word{\hat a}, \word{b}, \word{c},\word{d}) = n\cdot \vec \h(A,B,C,D) + O(\log n).
$$
(The hat in the notation of $\word{\hat a}$ is explained below; this string plays a distinguished role in the proof.)

The constructed quadruple $(\word{\hat a},\word{b},\word{c},\word{d})$ satisfies most requirements of the theorem:
$$
\info(\word{c}\lon \word{d}) \gg \info(\word{c}\lon \word{d}|\word{\hat a}) + \info(\word{c}\lon \word{d}|\word{b})+\info(\word{\hat a}\lon \word{b}),
$$
and $\info(\word{\hat a}\lon \word{b}|\word{c})=O(\log n)$. Only one requirement of the theorem is not satisfied:
the mutual information  $\info(\word{a}\lon \word{b})$ is much greater than
$\log n$. Thus, we need to transform  $(\word{\hat a},\word{b},\word{c},\word{d})$ so that (i) we keep the property 
$\square_{\word{\hat a}\word{b},\word{c}\word{d}}\ll 0$, (ii) $I(\word{\hat a}\lon \word{b}|\word{c})$ \emph{remains} logarithmic, 
and  (iii) $I(\word{\hat a}\lon \word{b})$ \emph{becomes} logarithmic. To this end, we modify string $\word{\hat a}$.

We apply Theorem~\ref{thm-muchnik}  for $\word{\hat a}$ and $\word{b}$ 
and get $\word{a}=Much(\word{\hat a}|\word{b})$ such that 
\begin{itemize}
\item $\C(\word{a}|\word{\hat a}) = O(\log n)$,
\item $\C(\word{a}) = \C(\word{\hat a}|\word{b}) + O(\log n)$,
\item $\C(\word{\hat a}|\word{a},\word{b}) = O(\log n)$.
\end{itemize}
Intuitively, string $\word{a}$ is the ``difference''  $\word{\hat a}$ \emph{minus} $b$ (in some sense, Muchnik's theorem  eliminates the mutual information between $\word{\hat a}$ and $b$). In what follows we show that the quadruple $(\word{a},\word{b},\word{c},\word{d})$ satisfies all the requirements of the theorem. 

It follows immediately from the definition of $\word{a}=Much(\word{\hat a}|\word{b})$ that
\begin{itemize}
\item $\C(\word{\hat a}|\word{a}) = \info(\word{\hat a}\lon \word{b}) + O(\log n)$,
\item $\info(\word{a}\lon \word{b}) = O(\log n)$,
\item $\info(\word{a}\lon \word{b}|\word{c})=O(\log n)$ (since for the original tuple we have $\info(\word{\hat a}\lon \word{b}|\word{c})=O(\log n)$).
\end{itemize}

The complexity profile of $(\word{a},\word{b},\word{c},\word{d})$ cannot be far different from
the complexity profile of $(\word{\hat a},\word{b},\word{c},\word{d})$. Indeed,  $C(\word{a}|\word{\hat a})=O(\log n)$
and $\C(\word{\hat a}|\word{a})=\info(\word{a}\lon \word{b}) + O(\log n)$. Hence, the difference
between  corresponding components of complexity profiles 
$\vec \C(\word{a},\word{b},\word{c},\word{d})$ and $\vec \C(\word{\hat a},\word{b},\word{c},\word{d})$ is at most 
$\info(\word{a}\lon \word{b}) + O(\log n) = O\left(n\cdot \frac{\log q}{q}\right)$. 

On the one hand, by the construction of  $\word{a}$ we have $\info(\word{a}\lon \word{b}) = O(\log n)$
and $\info(\word{a}\lon \word{b}|\word{c})=O(\log n)$. On the other hand, by the construction of the distribution $(A,B,C,D)_q$ we have
$$
\begin{array}{rcl}
\info(\word{c}\lon \word{d})  &=& n \cdot (1-\frac1q) + O(\log n),\\
\info(\word{c}\lon \word{d}|\word{b}) &=& n\cdot O\left(\frac{\log q}{q}\right)+ O(\log n),
\end{array}
$$
and
\begin{align*}
\info(\word{c}\lon \word{d}|\word{a}) &\le \info(\word{c}\lon \word{d}|\word{\hat a})+ \C(\word{\hat a}|\word{a}) = 
O\left( n\log\left(\frac{\log q}{q}\right) \right)+ O(\log n).
\end{align*}
Thus, for large enough $q$ we get
$
\info(\word{c}\lon \word{d})   -   \info(\word{c}\lon \word{d}|\word{a}) - \info(\word{c}\lon \word{d}|\word{b})  = \Omega(n).
$
\end{IEEEproof}

\begin{IEEEproof}[Proof of Theorem~\ref{th-counterexample-kolm}(b)]
We again use the distribution $(A,B,C,D)_q$
defined in Section~\ref{geometric-counterexample} (the value of $q$ is chosen later).
Let us denote this distribution $(A,B,C,D)_q$.
We apply to this distribution Lemma~\ref{lemma-fromh-to-c} and
obtain a quadruple of strings $(\word{\hat {\mathstrut a}},\word{\hat {\mathstrut b}},\word{\hat  {\mathstrut c}},\word{\hat  {\mathstrut d}})$ such that 
$$
\vec \C(\word{\hat {\mathstrut a}},\word{\hat {\mathstrut b}},\word{\hat {\mathstrut c}},\word{\hat {\mathstrut d}}) = n\cdot \vec \h(A,B,C,D) + O(\log n)
$$
(in what follows we will choose  $n$ large enough).
For the constructed $\word{\hat {\mathstrut a}},\word{\hat {\mathstrut b}},\word{\hat {\mathstrut c}},\word{\hat {\mathstrut d}}$ 
$$
\square_{\word{\hat {\mathstrut a}}\word{\hat {\mathstrut b}},\word{\hat {\mathstrut c}}\word{\hat {\mathstrut d}}} < - \kappa\cdot  n
$$
(for some $\kappa >0$)
and $\info(\word{\hat {\mathstrut a}}\lon \word{\hat{\mathstrut b}}|\word{\hat {\mathstrut c}})=O(\log n)$. However, this quadruple of strings does not 
satisfy the requirements of the theorem: the value of $\C(\word{\hat {\mathstrut c}}|\word{\hat {\mathstrut a}},\word{\hat {\mathstrut b}})$ is much greater than
$\log n$. We need to to modify  $\word{\hat {\mathstrut a}},\word{\hat {\mathstrut b}},\word{\hat {\mathstrut c}},\word{\hat {\mathstrut d}}$ so that (i) we keep the property
$\square_{\word{\hat {\mathstrut a}}\word{\hat {\mathstrut b}},\word{\hat {\mathstrut c}}\word{\hat {\mathstrut d}}}\ll0$, (ii) $\info(\word{\hat {\mathstrut a}}\lon \word{\hat{\mathstrut b}}|\word{\hat{\mathstrut c}})$ \emph{remains} logarithmic, 
and  (iii) $\C(\word{\hat{\mathstrut c}}|\word{\hat{\mathstrut a}},\word{\hat{\mathstrut b}})$ \emph{becomes} logarithmic.

We apply Theorem~\ref{thm-muchnik} to the constructed strings $\word{\hat{\mathstrut a}},\word{\hat{\mathstrut b}},\word{\hat{\mathstrut c}}$
and get $x=Much(\word{\hat{\mathstrut c}}|\word{\hat{\mathstrut a}},\word{\hat{\mathstrut b}})$ such that 
\begin{itemize}
\item $\C(\word{x}|\word{\hat c}) = O(\log n)$,
\item $\C(\word{x}) = \C(\word{\hat\mathstrut{ c}}|\word{\hat{\mathstrut a}},\word{\hat{\mathstrut b}}) + O(\log n)$,
\item $\C(\word{x}|\word{\hat{\mathstrut a}},\word{\hat{\mathstrut b}},\word{\hat{\mathstrut x}}) = O(\log n)$.
\end{itemize}
Let us consider the conditional complexity profile $\vec \C(\word{\hat {\mathstrut a}},\word{\hat {\mathstrut b}},\word{\hat {\mathstrut c}},\word{\hat {\mathstrut d}}|\word{x})$.
It is not hard to check that the components of 
$\vec \C(\word{\hat {\mathstrut a}},\word{\hat {\mathstrut b}},\word{\hat {\mathstrut c}},\word{\hat {\mathstrut d}}|\word{x})$ differ from the corresponding components of
$\vec \C(\word{\hat {\mathstrut a}},\word{\hat {\mathstrut b}},\word{\hat {\mathstrut c}},\word{\hat {\mathstrut d}})$ by at most 
$
\C(\word{x})+O(\log n).
$ 
Moreover,  we have $\info(\word{\hat{\mathstrut a}}\lon \word{\hat{\mathstrut b}}|\word{\hat{\mathstrut c}},\word{x})=O(\log n)$ and $\C(\word{\hat{\mathstrut c}}|\word{\hat{\mathstrut a}},\word{\hat{\mathstrut b}},\word{\mathstrut x})=O(\log n)$. This is exactly the complexity profile that we want to obtain (without relativization).
Thus, we would like to find a quadruple of strings whose complexity profile is close to the conditional  complexity profile  $\vec \C(\word{\hat {\mathstrut a}},\word{\hat {\mathstrut b}},\word{\hat {\mathstrut c}},\word{\hat {\mathstrut d}}|\word{x})$.
To this end, we apply the following lemma.
\begin{lemma}\label{lemma-cond-to-uncond-profile}
For a string $\word{x}$ and an $m$-tuple of strings $\mathcal{Y}=(\word{y}_1,\ldots,\word{y}_m)$ 
there exists an $m$-tuple $\mathcal{Z}=(\word{z}_1,\ldots,\word{z}_m)$ such that 
$$
\vec \C(\mathcal{Z}) = \vec \C(\mathcal{Y}|\word{x}) + O(\log N),
$$
where $N=\C(\word{y}_1,\ldots,\word{y}_m)$.
\end{lemma}
(See the proof in Appendix.)
From Lemma~\ref{lemma-cond-to-uncond-profile} it follows that 
there exists another tuple of strings $(\word{a},\word{b},\word{c},\word{d})$ such that
$$
\vec C(\word{a},\word{b},\word{c},\word{d}) = \vec C(\word{\hat {\mathstrut a}},\word{\hat {\mathstrut b}},\word{\hat {\mathstrut c}},\word{\hat {\mathstrut d}}|\word{x}) + O(\log n).
$$

On one hand, for the constructed tuple $\C(\word{c}|\word{a},\word{b}) = O(\log n)$ and $\info(\word{a}\lon \word{b}|\word{c})=O(\log n)$.
On the other hand,
$$
\begin{array}{ccl}
\info(\word{c}\lon \word{d})  &=& n- O\left( n\log\left(\frac{\log q}{q}\right) +\log n \right),\\
\info(\word{a}\lon \word{b}) &=& O\left( n\log\left(\frac{\log q}{q}\right) +\log n \right),\\
\info(\word{c}\lon \word{d}|\word{b}) &=& O\left( n\log\left(\frac{\log q}{q}\right) +\log n \right),\\
\info (\word{c}\lon \word{d}|\word{a})  &=&
 O\left( n\log\left(\frac{\log q}{q}\right) +\log n \right).
\end{array}
$$
Hence, for large enough $q$ we get
$$
\info(\word{c}\lon \word{d})   -  \info(\word{c}\lon \word{d}|\word{a}) - \info(\word{c}\lon \word{d}|\word{b}) - \info(\word{a}\lon \word{b}) = \Omega(n).
$$
\end{IEEEproof}

\section{Conclusion}

In this paper we discussed several conditional linear inequalities for Shannon entropy
and proved that these inequalities are essentially conditional, \ie they cannot be obtained
by restricting an unconditional linear inequality to the corresponding subspace. 
We proved that there are two
types of essentially conditional inequalities: some of them hold for all almost entropic points,
while the other are valid only for entropic points. We discussed the geometric meaning of 
inequalities of the first type --- every essentially conditional inequality for almost 
entropic points corresponds to an infinite family of unconditional linear inequalities, and 
implies the fact that the cone of almost entropic points is not polyhedral.
We also proved that some (but not all) essentially conditional inequalities can be translated
in the language of Kolmogorov complexity.

Many questions remain unsolved:

\begin{itemize} 
\item Does Inequality ($\II$) hold for almost entropic points?
\item All known essentially conditional inequalities have at least two linear constraints. Do there 
exist any essentially conditional inequalities with a constraint of co-dimension~$1$?
\item For the proven inequalities $(\IV'\mbox{-Kolm})$  and $(\IV'\mbox{-Kolm})$, 
if the constraints hold up to $O(\log n)$, then the inequality holds up to $O(\sqrt{n\log n})$.
Do there exists any essentially conditional inequality for Kolmogorov complexity 
with $O(\log n)$-terms both in the constraints and in the resulting inequality? 
\item The  essentially conditional linear inequalities that do not hold for 
almost entropic points remain unexplained. Do they have any geometric, physical meaning?
\item Mat\'{u}\v{s} proved that for $n\ge4$ random variables
the  cone of all almost entropic points is not polyhedral, \ie cannot be represented as 
an intersection of a finite number of half-spaces.
Can we represent this  cone  as an intersection of \emph{countably}
many tangent half-spaces? 
\item It would be interesting to 
establish the connection between (i)~conditional information inequalities that hold  for almost entropic
points, (ii)~infinite families of unconditional linear inequalities, and (iii)~non linear information inequalities.
(In~\cite{nonlinear} a family of linear inequalities from \cite{Matus-inf}
was converted into a quadratic inequality for entropies. This result seems to be an instance of a more general
phenomenon, but it is difficult to make a more precise conjecture.)
\end{itemize}

\section*{Acknowledgments}

This paper was motivated by a series of works by Franti\v{s}ek Mat\'{u}\v{s}.
We thank him also for his personal comments on \cite{condineq}
(especially for pointing out a mistake in that paper). 
The authors  are indebted to Alexander Shen and Ivan Sebedash
for prolific discussions. We  would like to thank all our colleagues in 
the Kolmogorov seminar in Moscow and the group Escape in Montpellier.


\section{Appendix}


\begin{IEEEproof}[Proof of lemma~\ref{lemma-iv-1}]
We say that two pairs of values $(c_1,d_1)$ and
$(c_2,d_2)$ are equivalent if the conditional distributions 
on $A$ given conditions $(C = {c}_1\ \&\ D=d_1)$ and
$(C = c_2\ \&\  D=d_2)$ are the same, \ie
for every $a$
 $$
 \prob[A = a | C = c_1, D=d_1]
 {=}
  \prob[A =a | C = c_2, D=d_2].
 $$
The class of equivalence corresponding to random value of $(C,D)$ is also a 
random variable. We denote this random variable by $W$.

From the definition of $W$ it follows that the conditional distribution of $A$ given  the triple $(C,D,W)$ 
is exactly the same as the conditional distribution of $A$ given $W$.
Hence, $I(A\lon C,D|W)=0$.

Further, from $I(A\lon D|C)=0$ it follows that $W$ functionally depends on $C$
and from $I(A\lon C|D)=0$ it follows that $W$ functionally depends on $D$
(the conditional distributions of $A$ given  $C= c_1$, or given
 $D=d_1$, or given $(C = c_1\ \&\  D=d_1)$
 are all the same). Hence, $\h(W | C) =\h(W|D)= 0$.
\end{IEEEproof}

\begin{IEEEproof}[Proof of Lemma~\ref{rel-lemma}]
First we serialize $(A,B,C,D)$,  \ie
we take $m$ i.i.d. copies of the initial distribution. 
The result is a distribution $(A^m,B^m,C^m,D^m)$ whose entropy profile is
equal to the entropy profile of $(A,B,C,D)$ multiplied by $m$. In particular, 
we have $I(A^m\lon B^m|C^m) = 0$.
Then, we apply Slepian\textendash{}Wolf coding (Lemma~\ref{lemma-sw})
and get a $Z=SW(C^m|A^m,B^m)$ such that 
\begin{itemize}
\item $\h(Z|C^m)=0$,
\item $\h(Z)=\h(C^m| A^m,B^m) + o(m)$,
\item $\h(C^m|A^m,B^m,Z^m)=o(m)$.
\end{itemize}
The  profile of  conditional entropies of $(A^m,B^m,C^m,D^m)$ given $Z$
differs from the entropy profile of $(A^m,B^m,C^m,D^m)$ by at most $\h(Z) = m\cdot
\h(C|A,B) + o(m)$ (\ie the differences between $\h(A^m)$ and $\h(A^m|Z)$, between $\h(B^m)$ and $\h(B^m|Z)$, 
etc., are not greater than $\h(Z)$).  Also, if in the original distribution $I(A\lon B|C) = 0$,
then $I(A^m\lon B^m|C^m, Z) = I(A^m\lon B^m|C^m) = 0$.

We would like to ``relativize'' $(A^m,B^m,C^m,D^m)$ conditional on $Z$ and get a new distribution
for a quadruple $(A',B',C',D')$ whose unconditional entropies are equal to the corresponding conditional
entropies of $(A^m,B^m,C^m,D^m)$ given $Z$. This relativization procedure is not straightforward since
for different values of $Z$, the corresponding  conditional distributions on $(A^m,B^m,C^m,D^m)$ 
can be  very different. 
The simplest way to overcome this obstacle is the method  
of quasi-uniform distributions proposed by Chan and Yeung in \cite{chan-yeung}.
\begin{definition}[Chan\textendash{}Yeung,\cite{chan-yeung}] 
A random variable $U$  distributed  on a finite set $\mathcal{U}$ is called \emph{quasi-uniform} 
if  the probability distribution function of $U$ is constant over its support (all values of $U$
have the same probability). That is, there exists a real $c>0$ such that  $\prob[U=u]\in \{0,c\}$
for all $u\in \mathcal{U}$.
A tuple of  random variables $(X_1,\ldots,X_n)$ is called quasi-uniform if for any non-empty
subset $\{i_1,\ldots,i_k\}\subset\{1,\ldots,n\}$ the joint distribution
$(X_{i_1},\ldots,X_{i_k})$ is quasi-uniform.
\end{definition}

In \cite[Theorem~3.1]{chan-yeung} it was proven that for every distribution $(X_1,\ldots,X_n)$ 
and for every $\delta>0$ there exists an integer $k$ and a quasi-uniform distribution $(\hat X_1,\ldots,\hat X_n)$ such that 
$
\| \vec \h(X_1,\ldots,X_n) - \frac1k \vec \h(\hat X_1,\ldots,\hat X_n)\|<\delta.
$
We apply this theorem to the distribution $(A^m,B^m,C^m,D^m,Z)$ and conclude that for
every $\delta>0$ there exists an integer $k$ and a quasi-uniform distribution $(\hat A,\hat B,\hat C,\hat D,\hat Z)$ such that 
$$
\| \vec \h(A^m,B^m,C^m,D^m,Z) - \frac1k \vec \h(\hat A,\hat B,\hat C,\hat D,\hat Z)\|<\delta.
$$
In a quasi-uniform distribution, the conditional distributions of $(\hat A,\hat B,\hat C,\hat D)$  under  the constraints  $\hat Z=z$   for different values $z$ are all isomorphic to each other. 
Hence, for every $z$ the entropies of  the conditional distributions of $\hat A$, $\hat B$, $(\hat A,\hat  B)$, etc.,  given $\hat Z=z$ are equal to $\h(\hat A|\hat Z)$, $\h(\hat B|\hat Z)$, $\h(\hat A, \hat B|\hat Z)$, etc., respectively. Thus, for a quasi-uniform distribution we can perform a relativization as follows.
We fix any value $z$ of $\hat Z$ and take the conditional distribution  on
$(\hat A,\hat B,\hat C,\hat D)$ given $\hat Z={z}$.
In this conditional distribution the entropy of 
$\hat C$ given $(\hat A,\hat B)$ is not greater than
\[ k\cdot (\h(C^m|A^m,B^m,Z) + \delta)=k\cdot (\delta +o(m)).\]
If $\delta$ is small enough, then 
all entropies of $(\hat A,\hat B,\hat C,\hat D)$ given $\hat Z=z$ differ from
the corresponding components of $km \cdot \vec \h(A,B,C,D)$  by at most 
$\h(\hat Z) \le km\cdot \h(C|A,B) + o(km)$.

Moreover, the mutual information between $\hat A$ and $\hat B$ given $(\hat C,\hat Z)$ is the same
as the mutual information between $\hat A$ and $\hat B$ given only $\hat C$, since $Z$ functionally depends on $C$.
If  in the original distribution $I(A\lon B|C) = 0$, then the mutual information
between  $(\hat A,\hat B)$ given $(\hat C,\hat Z)$ is $o(km)$.

We choose $\delta$ small enough (e.g., $\delta=1/m$)
and let $(A',B',C',D')$ be the constructed above conditional distribution.
\end{IEEEproof}

\begin{IEEEproof}[Proof of lemma~\ref{lemma-cond-to-uncond-profile}]
We are given a tuple of strings $\mathcal{Y}=(\word{y}_1,\ldots,\word{y}_m)$. For every subset
of indices $W=\{i_1,\ldots,i_k\}$ we denote 
$\mathcal{Y}_W = (\word{y}_{i_1},\ldots,\word{y}_{i_k})$, assuming $1\le i_1<\cdots<i_k\le m$.

Let $S$ be the set of all tuples $\mathcal{Y}'=(\word{y}'_1,\ldots,\word{y}'_m)$ such that 
for all  $U,V\subset \{1,\ldots,m\}$
$$
\C(\mathcal{Y}'_U|\word{x}) \le \C(\mathcal{Y}_U|\word{x})
\mbox{ and }
\C(\mathcal{Y}'_U|\mathcal{Y}'_V,\word{x}) \le \C(\mathcal{Y}_U|\mathcal{Y}_V,\word{x}).
$$
In particular, this set contains the initial tuple $\mathcal{Y}$. 
Further, for each $U\subset \{1,\ldots,m\}$
we denote 
$$
h_U = \left\lceil \log \left[\begin{array}{l}%
\mbox{the size of the projection}\\
\mbox{of $S$ onto $U$-coordinates}
\end{array}
\right] \right\rceil
$$
and
$$
h_U^{\bot} = \left\lceil \log \left[%
\begin{array}{l}
\mbox{the maximal size of a section of $S$}\\
\mbox{for some fixed $U$-coordinates}
\end{array}
\right] \right\rceil.
$$
E.g., if $U=\{1\}$, then $h_{\{1\}}$ is equal to the number of all strings $\word{y}_1'$ such that for
some strings  $\word{y}_2',\ldots,\word{y}_m'$ the tuple  $(\word{y}_1',\word{y}_2',\ldots,\word{y}_m')$ belongs to $S$. 
Similarly, $h_{\{1\}}^\bot$ is by definition equal to 
 $$
 \max_{\word{y}_1'}\  \# \{  (\word{y}_2',\ldots,\word{y}_m') 
   \mbox{ such that }(\word{y}_1',\word{y}_2',\ldots,\word{y}_m')\in S \}.
 $$
 
The cardinality of $S$ is less than $2^{C(\mathcal{Y}|\word{x})+1}$ since for
each tuple $\mathcal{Y}'$ in $S$ there exists a program of length
at most $\C(\mathcal{Y}|\word{x})$ which translates $\word{x}$ to $\mathcal{Y}'$.
Similarly, for each $U$
$$
h_U \le \C(\mathcal{Y}_U)+1 \mbox{ and } 
 h_U^{\bot}  \le \C(\mathcal{Y}_{\bar U}|\mathcal{Y}_{U})+1,
$$ 
where $\bar U = \{1,\ldots,m\}\setminus U$.
On the other hand, the cardinality of $S$ cannot be less than $2^{ C(\mathcal{Y}|x)-O(\log N)}$,
since we can specify $\mathcal{Y}$ given $x$ by the list of numbers $h_U$ and  $h_U^{\bot}$ 
and the ordinal number of $\mathcal{Y}$ in the standard enumeration of $S$.

We need only $O(\log N)$ bits to specify all numbers $h_U$ and  $h_U^{\bot}$ (the constant
in this $O(\log N)$ term depends on  $m$  but not on $N$).
Given all  $h_U$ and  $h_U^{\bot}$ we can find some set $S'$ such that
the sizes of all projections and maximal sections of $S'$ are equal to the corresponding
$h_U$ and  $h_U^{\bot}$.  At least one such set exists (e.g., set  $S$ satisfies all these conditions), so we
can find one such set by brute-force search. Note that we do not need to
know $\word{x}$ to run this search.

For each tuple $\mathcal{Z}=(\word{z}_1,\ldots,\word{z}_m)$ in $S'$ we have 
$$
\C(\mathcal{Z}) \le h_{\{1,\ldots,m\}}+O(\log N)= \C(\mathcal{Y}| \word{x})+O(\log N)
$$
since we can specify this tuple by the list of all $h_U$ and  $h_U^{\bot}$
and the index of $\mathcal{Z}$ in the list of elements $S'$.
Similarly, for each set of indices $U$
\begin{equation}\label{eq-star3}
\C(\mathcal{Z}_U) \le h_U+O(\log N) = \C(\mathcal{Y}_U| \word{x})+O(\log N)
\end{equation}
and
$$
\C(\mathcal{Z}_{\bar U}|\mathcal{Z}_{ U}) \le h_U^\bot +O(\log N)= \C(\mathcal{Y}_{\bar U}|\mathcal{Y}_{U},\word{x})+O(\log N).
$$

Let $\mathcal{Z}=(\word{z}_1,\ldots,\word{z}_m)$ be some tuple in $S'$ with maximal possible complexity.
Then $\C(\mathcal{Z})  = \C(\mathcal{Y}| \word{x})+O(\log N) $ (complexity of $\mathcal{Z}$ cannot
be much less since the cardinality of $S'$ is equal to $2^{ \C(\mathcal{Y}|x)-O(\log N)}$). For this tuple $\mathcal{Z}$,
Inequality~(\ref{eq-star3}) becomes an equality
 \begin{equation*}
\C(\mathcal{Z}_U)  =  \C(\mathcal{Y}_U| \word{x})+O(\log N)
 \end{equation*}
for all tuples $U$.
 Indeed, if $\C(\mathcal{Z}_U) $ is much less than $h_U$ for some $U$, then
$$
\C(\mathcal{Z})  = \C(\mathcal{Z}_U)  + \C(\mathcal{Z}_{\bar U}| \mathcal{Z}_U)  + O(\log N)
$$
is much less than
$$
\C(\mathcal{Y}|\word{x})  = \C(\mathcal{Y}_U|x)  + \C(\mathcal{Y}_{\bar U} |\mathcal{Y}_{U},\word{x})  + O(\log N),
$$
and we get a contradiction with the choice of  $\mathcal{Z}$. Hence, the difference
between the corresponding components of complexity profiles
$\vec \C(\mathcal{Z})$ and  $\vec \C(\mathcal{Y}|\word{x})$ is bounded by $O(\log N)$.
\end{IEEEproof}

\end{document}